\documentclass{IOS-Book-Article}
\usepackage[utf8]{inputenc}
\usepackage{mathptmx}
\usepackage{graphicx}
\usepackage{ltl}
\usepackage{csquotes}
\usepackage{amsmath}
\usepackage{xcolor}
\usepackage{mathtools}
\usepackage{tikz}
\usepackage{amsmath} 
\usepackage{amssymb}
\usepackage{wasysym}
\usepackage{mathtools}
\usepackage{lmodern}
\usepackage{mathrsfs}
\usepackage{float}
\usepackage{colortbl}
\usepackage{multirow}
\newcommand\tuple[1]{\langle #1 \rangle}

\usetikzlibrary
  { arrows
  , automata
  , positioning
  , calc
  }

\colorlet{darkred}{red!80!black}
\colorlet{darkblue}{blue!80!black}
\colorlet{darkgreen}{green!80!black}

\colorlet{darkyellow}        {yellow!60!black}

\newcommand\comment[1]{}

\newcommand{\size}[1]{\ensuremath{|#1|}}

\newcommand{\nats}{\ensuremath{\mathbb{N}}}

\newcommand{\set}[1]{\ensuremath{\{#1\}}}

\renewcommand{\emptyset}{\varnothing}

\newcommand{\lang}{\ensuremath{\mathcal{L}}}

\newcommand{\aut}{\ensuremath{\mathcal{A}}} 

\newcommand{\tree}{\ensuremath{\mathcal{T}}}

\newcommand\produ{\times}

\newcommand{\inputs}{\ensuremath{\mathcal{I}}}
\newcommand{\outputs}{\ensuremath{\mathcal{O}}}
\newcommand{\mealy}{\ensuremath{\mathcal{M}}}

\tikzset{
  node distance=6em,
  >=stealth',
  pl0/.style={draw,circle,minimum size=9mm,fill=blue!20},
  pl1/.style={draw,minimum size=9mm,fill=red!30},
  ac0/.style={draw,circle,minimum size=8mm,fill=blue!20,double=blue!20},
  ac1/.style={draw,minimum size=8mm,fill=red!30,double=red!30},
  r/.style={bend left=18},
}

\tikzstyle{state} = [draw,circle,minimum size=9mm, fill=blue!20]
\tikzstyle{accepting} = [double=blue!20]
\tikzstyle{marked} = [fill=red!40,double=red!40]
\tikzstyle{win} = [scale=5,yellow!80!black,fill=yellow!50,rounded corners=8,draw]
\tikzstyle{strat} = [ultra thick, blue]
\tikzset{initial text ={}}

\tikzset{
    between/.style args={#1 and #2}{
         at = ($(#1)!0.5!(#2)$)
    }
}

\newcommand\AP{\mathit{AP}}
\newcommand\request{\mathit{R}}
\newcommand\grant{\mathit{G}}
\newtheorem{example}{Example}

\newtheorem{lemma}{Lemma}

\newcommand{\ttrans}[3]{\ensuremath{\textsc{trans}(#1,#2,#3)}} 
\newcommand{\rgstate}[2]{\ensuremath{\textsc{rgstate}(#1,#2)}}
\newcommand{\tlabel}[3]{\ensuremath{\textsc{output}(#1,#2,#3)}}  
\newcommand{\annotationd}[3]{\ensuremath{\textsc{annotation}(#1,#2,#3)}}  
\newcommand{\annotation}[2]{\ensuremath{\textsc{annotation}(#1,#2)}}  
\newcommand{\edge}[2]{\ensuremath{\textsc{edge}(#1,#2)}}  
\newcommand{\redge}[2]{\ensuremath{\textsc{redge}(#1,#2)}}  
\newcommand{\bedge}[2]{\ensuremath{\textsc{bedge}(#1,#2)}}  
\newcommand{\wtreed}[2]{\ensuremath{\textsc{wtree}(#1,#2)}}  
\newcommand{\wtree}[1]{\ensuremath{\textsc{wtree}(#1)}}  
\newcommand{\rboundd}[2]{\ensuremath{\textsc{rbound}(#1,#2)}}  
\newcommand{\rbound}[1]{\ensuremath{\textsc{rbound}(#1)}}

\newcommand{\allowed}[2]{\ensuremath{\textsc{visited}(#1,#2)}}  
\newcommand{\sccd}[3]{\ensuremath{\textsc{scc}(#1,#2,#3)}}  
\newcommand{\scc}[2]{\ensuremath{\textsc{scc}(#1,#2)}}  
\newcommand{\curlyF}{\ensuremath{\mathcal{F}}}

\newtheorem{theorem}{Theorem}

\begin{document}

\begin{frontmatter}              

\title{Reactive Synthesis: Towards Output-Sensitive Algorithms}
\runningtitle{Reactive Synthesis}

\author{Bernd Finkbeiner \and Felix Klein}

\runningauthor{B. Finkbeiner, F. Klein}
\address{Universit\"at des Saarlandes}

\begin{abstract}
  Reactive synthesis is a technology for the automatic construction
  of reactive systems from logical specifications.
  In these lecture notes, we study different
  algorithms for the reactive synthesis problem of linear-time
  temporal logic (LTL).  The classic game-based synthesis algorithm is
  input-sensitive in the sense that its performance is asymptotically
  optimal in the size of the specification, but it produces
  implementations that may be larger than necessary.  We contrast this
  algorithm with output-sensitive algorithms for reactive synthesis,
  i.e., algorithms that are optimized towards the size or structural
  complexity of the synthesized system. We study the bounded synthesis
  algorithm, which produces an implementation with a minimal number of
  states, and the bounded cycle synthesis algorithm, which
  additionally guarantees that the number of cycles of the
  implementation is minimal.
 \end{abstract}

\end{frontmatter}

\thispagestyle{empty}
\pagestyle{empty}

\section{Introduction}

Hardware circuits, communication protocols, and embedded controllers
are typical examples of \emph{reactive systems}~\cite{101990}, i.e., computer
systems that maintain a continuous interaction with their
environment. Reactive systems play a crucial role in many applications
in transport systems, building technology, energy management, health
care, infrastructure, and environmental protection.  Designing
reactive systems is difficult, because one needs to anticipate every possible behavior of the environment and prepare an appropriate response.

Synthesis is a technology that constructs reactive systems 
\emph{automatically} from a
logical specification: that is, after the specification of the system is complete, no
further manual implementation steps are necessary. The developer focuses on ``what''
the system should do instead of ``how'' it should be done. Because
synthesis analyzes objectives, not implementations, it can be applied
at an early design stage, long before the system has been
implemented. The vision is that a designer analyzes the design
objectives with a synthesis tool, automatically identifies competing
or contradictory requirements and obtains an error-free prototype
implementation. Coding and testing, the most expensive stages of
development, are eliminated from the development process.

The automatic synthesis of implementations from specifications is one
of the grand challenges of computer science. Its pursuit dates back at
least to Alonzo Church~\cite{Church/57/Applications} and has ignited
research on many fundamental topics, notably on the connection between
logics and automata, on algorithmic solutions of infinite games over
finite graphs~\cite{Buchi1969}, and on the theory of automata over
infinite objects~\cite{Rabin/72/Automata}. It is only in the
last decade, however, that the theoretical ideas have been translated
into practical tools~(cf.~\cite{Jobstm07c,DBLP:conf/tacas/Ehlers11,DBLP:conf/cav/BohyBFJR12,BGHKK12,Faymonville2017b}).
The tools have made it possible to tackle real-world design problems, such as the synthesis of an arbiter for the  \emph{AMBA AHB bus}, an open industrial standard for the on-chip communication and management of functional blocks in system-on-a-chip (SoC) designs~\cite{Bloem+others/07/Automatic}.

A common argument \emph{against} synthesis is its complexity. It is natural to compare synthesis with the verification problem, where the implementation is already given, and
one needs to check whether the specification is satisfied.  For both
synthesis and verification, the most commonly used specification
language is linear-time temporal logic (LTL).  Measured in the size of
an LTL specification, the synthesis of a single-process finite-state
machine is 2EXPTIME-complete, while the corresponding verification
problem is in PSPACE.
But is this
comparison between verification and synthesis fair? The high
complexity of synthesis is due to the fact that there exist small
LTL formulas that can only be realized by very large implementations. 
As a result, synthesis
``looks'' much more expensive than verification, because the size of the implementation is an explicit
parameter in the complexity of verification, and left implicit in the
complexity of synthesis.

This paper gives an introduction to a new class of synthesis algorithms, whose
performance is measured not only in the size of the specification, i.e., the
input to the synthesis algorithm, but also in the size and complexity of
the implementation, i.e., the output of the synthesis algorithm. Such
algorithms are called \emph{output sensitive}.
The prototypical output-sensitive synthesis approach is \emph{bounded synthesis}.
In bounded synthesis, we look for an implementation where the number of states
is limited by a given bound. By incrementally increasing the bound, bounded
synthesis can be used to find a minimal implementation.

We first describe the classic game-theoretic approach
to synthesis in Section~\ref{sec:games}, and then the bounded synthesis
approach in Section~\ref{sec:boundedsynthesis}.
The two approaches differ fundamentally. The game-based approach is to translate the given LTL formula
into an equivalent deterministic automaton, and then use the state
space of the deterministic automaton to define a two-player game.
In this game, the ``output player'' sets the outputs of the system
and attempts to satisfy the specification, i.e., ensures that the
resulting play is accepted by the automaton, and the ``input player''
sets the inputs and attempts to ensure that the play violates the
specification, i.e., is rejected by the automaton.
This game can be solved automatically, and a winning strategy for
the output player can, if it exits, be translated into an
implementation that is guaranteed to satisfy the specification.
Unfortunately, the translation from LTL to deterministic automata
is doubly exponential, which results in the 2EXPTIME complexity.
In bounded synthesis, the LTL formula is not translated to a
deterministic automaton; instead, its negation is translated to a nondeterministic automaton.
This translation is single, rather than double exponential.
The nondeterministic automaton suffices to check if a given
implementation is correct: the implementation is correct if its
product with the automaton does not contain an accepting path. In bounded synthesis, we ``guess''
an implementation of bounded size and make sure it is correct. This is
done via propositional constraint solving: we build a constraint system
that is satisfiable if and only if an implementation that is correct with
respect to the automaton.



The reduction of the synthesis problem to a constraint solving problem
opens the possibility to add further constraints in order to focus the
search towards the most desirable solutions.  In
Section~\ref{sec:boundedcyclesynthesis}, we describe such an
extension: \emph{bounded cycle synthesis}.  In addition to the number of
states, bounded cycle synthesis also bounds the number of cycles in
the implementation. This leads to implementations that are not only
small but also structurally simple.

\section{The Synthesis Problem}

In reactive synthesis, we transform a temporal specification into an implementation that is guaranteed to satisfy the specification for all possible inputs of the environment. In the following, we consider formulas of linear-time temporal logic (LTL) over a set of atomic propositions $\AP = I \dot\cup O$ that is partitioned into a set of \emph{inputs} $I$ and a set of \emph{outputs} $O$.
A \emph{trace} $t$ is an infinite sequence over subsets of the atomic propositions. We define the set of traces $\mathit{TR} \coloneqq (2^\mathit{AP})^\omega$.
An LTL formula describes a subset of $\mathit{TR}$. The idea is that in each step of a computation, the inputs are chosen by the environment, and the outputs are chosen by the system under construction. In a correctly synthesized system, all possible sequences satisfy the LTL formula.

\paragraph{\bf Linear-time temporal logic (LTL).}
\label{sec:ltl}
Linear-time temporal logic (LTL)~\cite{DBLP:conf/focs/Pnueli77} combines the usual Boolean connectives with temporal modalities such as the \emph{Next} operator $\LTLnext$ and the \emph{Until} operator $\LTLuntil$. The syntax of LTL is given by the following grammar:
\begin{align*}
\varphi~& \Coloneqq~p~~|~~\neg \varphi~~|~~\varphi \vee \varphi~~|~~\LTLnext \varphi~~|~~\varphi\, \LTLuntil \varphi
\end{align*}
where $p \in \mathit{AP}$ is an atomic proposition.
$\LTLnext \varphi$ means that $\varphi$ holds in the \emph{next} position of a trace; $\varphi_1 \LTLuntil \varphi_2$ means that $\varphi_1$ holds \emph{until} $\varphi_2$ holds.
There are several derived operators, such as
$\LTLdiamond \varphi \equiv \mathit{true}\, \LTLuntil \varphi$,
$\LTLsquare \varphi \equiv \neg \LTLdiamond \neg \varphi$, and $\varphi_1\,\mathcal\, \mathcal W\, \varphi_2 \equiv (\varphi_1 \LTLuntil \varphi_2) \vee \LTLsquare \varphi_1$.  
$\LTLdiamond \varphi$ states that $\varphi$ will \emph{eventually} hold in the future and $\LTLsquare \varphi$ states that $\varphi$ holds \emph{globally}; $\mathcal W$ is the \emph{weak} version of the \emph{until} operator. 

We use the following notation to manipulate traces:
let $t \in \mathit{TR}$ be a trace and $i \in \mathbb{N}$ be a natural number. $t[i]$ denotes the $i$-th element of $t$. Therefore, $t[0]$ represents the starting element of the trace. Let $j \in \mathbb{N}$ and $j \geq i$, then $t[i,j]$ denotes the sequence $t[i]~t[i+1]\ldots t[j-1]~t[j]$, and $t[i, \infty]$ denotes the infinite suffix of $t$ starting at position $i$. Let $p \in \mathit{AP}$ and $t \in \mathit{TR}$.
The semantics of an LTL formula is defined as the smallest relation $\models$ that satisfies the following conditions:
\begin{align*}
& t \models p &&~\text{iff} \hspace{5ex} p \in t[0] \\
& t \models \neg \psi &&~\text{iff} \hspace{5ex} t \not \models \psi \\
& t \models \psi_1 \vee \psi_2 &&~\text{iff} \hspace{5ex} t \models \psi_1~\text{or}~t \models \psi_2 \\
& t \models \LTLnext \psi &&~\text{iff} \hspace{5ex} t [1,\infty] \models \psi \\
& t \models \psi_1 \LTLuntil \psi_2 &&~\text{iff}\hspace{5ex} \text{there exists}~i \geq 0 : t[i,\infty] \models \psi_2 \\
& &&\hspace{7.8ex} \text{and for all}~0 \leq j < i~\text{we have}~t[j,\infty] \models \psi_1
\end{align*}

\begin{example}
  \label{example:arbiter}
Suppose, for example, we are interested in constructing an
arbiter circuit. Arbiters are used when more than one client 
needs access to some shared resource, such as a communication bus. To
access the resource, the client sends a \emph{request} signal
$\request$ and waits until it receives a \emph{grant} signal $\grant$
from the arbiter. The task of the arbiter is to answer each request
with a grant without giving grants to the two clients at the same
time. In LTL, an arbiter with two clients can be specified as a
conjunction of three properties:
\[
\begin{array}{lr}
\LTLsquare\, (\neg \grant_1 \vee \neg \grant_2) & \qquad \qquad \mbox{(mutual exclusion)}\\
\LTLsquare\, (\request_1 \rightarrow \LTLdiamond \grant_1) & \mbox{(response 1)}\\
\LTLsquare\, (\request_2 \rightarrow \LTLdiamond \grant_2) & \mbox{(response 2)}\\
\end{array}
 \]
The \emph{mutual exclusion} property states that at every point in
time $x$, at most one grant signal can be set; the \emph{response}
properties state that if a request is made at some point in time, then
there must exist a point in time, either immediately or later, where the
corresponding grant signal is set.
\end{example}

\paragraph{\bf Implementations.}
We represent the result of the synthesis process as a finite-state machine.
Let the set $\AP= I \dot\cup O$ of atomic propositions be, as before, partitioned into the inputs $I$ and the outputs $O$.
A \emph{Mealy machine} over $I$ and $O$  has the form
$ \mealy = (S, s_{0}, \delta, \gamma) $ where $S$ is a finite
set of states, $ s_{0} \in S $ is the initial state,
$\delta \colon S \times 2^I \rightarrow S $ is the transition
function, and $\gamma \colon S \times 2^I \rightarrow 2^O$
is the output function. The output of the Mealy machine thus depends on the
current state and the last input letter. A \emph{path}
of a Mealy machine is an infinite sequence
$ p = (s_{0},\sigma_{0})(s_{1},\sigma_{1})(s_{2},\sigma_{2}) \ldots
\in (S \times 2^{\AP})^{\omega} $ of states and sets of atomic propositions that starts with the
initial state $s_0$ and where
$ \delta(s_{n},I \cap \sigma_{n}) = s_{n+1} $ and
$ \gamma(t_{n},I \cap \sigma_{n}) = O \cap \sigma_{n} $
for all $ n \in \mathbb N $.
We refer to the projection of a path $p$ to its second component 
$ \pi = \sigma_{0}\sigma_{1}\sigma_{2}\ldots \in
\Sigma^{\omega} $,
as a \emph{computation} of the Mealy machine.
The Mealy machine \emph{satisfies} the LTL
formula $\varphi$, denoted by $M \models \varphi$, if all its computations satisfy~$\varphi$.

\begin{figure}[t]
\begin{tikzpicture}
  \node at (6,-0.4) {
    \scalebox{0.7} {
      \begin{tikzpicture}[auto,semithick]
        \node[state, initial] (A)                         {$ t_{0} $};
        \node[state]          (B) [right of=A,xshift=8em] {$ t_{1} $};

        \path[->]
        (A) edge[loop above] node[above] {
              $ \emptyset, \set{ \request_{1} } \to \set{ \grant_{1} } $
            } (A)
        (A) edge[bend left=10] node[above] {
              $ \set{ \request_{2} }, \set{ \request_{1}, \request_{2} } \to \set{ \grant_{1} } $
            } (B)
        (B) edge[loop above] node[above] {
              $ \emptyset, \set{ \request_{2} } \to \set{ \grant_{2} } $
            } (B)
        (B) edge[bend left=10] node[below] {
              $ \set{ \request_{1} }, \set{ \request_{1}, \request_{2} } \to \set{ \grant_{2} } $
            } (A)
        ;
      \end{tikzpicture}
    }
  };

  \node at (0,0) {
    \scalebox{0.7} {
      \begin{tikzpicture}[auto,semithick]
        \node[state, initial] (A)                                      {$ s_{0} $};
        \node[state]          (B) [below of=A,xshift=-8em,yshift=-5em] {$ s_{1} $};
        \node[state]          (C) [below of=A,xshift=8em,yshift=-5em]  {$ s_{2} $};

        \path[->]
        (A) edge[loop above] node[above] {
              \begin{tabular}{c}
                $ \emptyset \to \emptyset $ \\ 
                $ \set{ \request_{1} } \to \set{ \grant_{1} } $ \\
                $ \set{ \request_{2} } \to \set{ \request_{2} } $
              \end{tabular} 
            } (A)
        (A) edge[out=0,in=75] node[below,rotate=-55,yshift=1.7em] {
              $ \set{ \request_{1}, \request_{2} } \to \set{ \grant_{1} } $
            } (C)
        (B) edge node[above,rotate=55] {
              $ \emptyset, \set{ \request_{1} } \to \set{ \grant_{1} } $
            } (A)
        (B) edge[bend left=10] node[above] {
              $ \set{ \request_{2} }, \set{ \request_{1}, \request_{2}} \rightarrow \set{ \grant_{1} } $
            } (C)
        (C) edge[bend left=10] node[below] {
              $ \set{ \request_{1} }, \set{ \request_{1}, \request_{2}} \rightarrow \set{ \grant_{2} } $
            } (B)   
        (C) edge node[above,rotate=-55] {
              $ \emptyset, \set{ \request_{2} } \to \set{ \grant_{2} } $
            } (A)
        ;
      \end{tikzpicture}
    }
  };
\end{tikzpicture}

\caption{Two Mealy machines implementing the arbiter specification. 
  \label{fig:mealymachines}}
\end{figure}
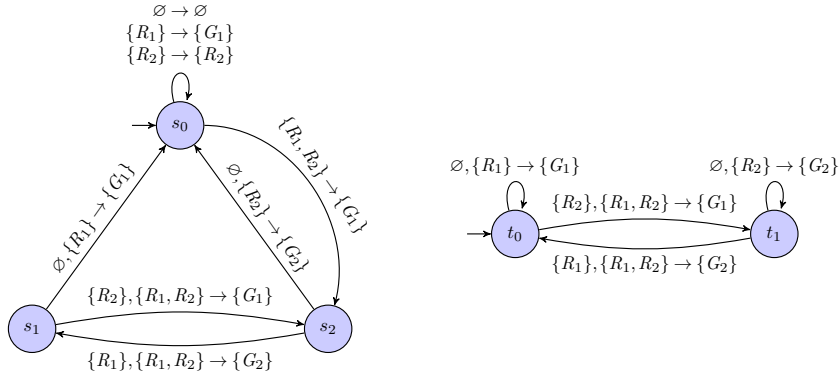

\begin{example}
Figure~\ref{fig:mealymachines} shows two Mealy machines that implement
the arbiter specification from Example~\ref{example:arbiter}. The Mealy machine shown on the left carefully answers every request and only issues a grant if there is an open request. The machine on the right always issues the grant to the same client, initially to the first client, and switches to the other client as soon as there is a request from the other client.
Both machines satisfy the specification from Example~\ref{example:arbiter}.
\end{example}

\paragraph{\bf Realizability and Synthesis.} We say that an LTL formula $\varphi$ is \emph{realizable} if there exists a Mealy machine~$ \mealy $ over the same inputs $ I $ and outputs $ O $ as $ \varphi $ such that $ \mealy \models \varphi$. The \emph{synthesis problem} of an LTL formula $\varphi$ is to determine whether $\varphi$ is realizable and, if the answer is yes, to construct a Mealy machine $ \mealy $ such that $ \mealy \models \varphi$.

\section{Model checking}

Before we address the synthesis problem, we take a quick detour into model checking. In model checking, the implementation is already given and we are interested in determining whether the implementation is correct. 

Given a Mealy machine~$\mealy$ and an LTL formula $\varphi$, model checking determines whether $\mealy$ satisfies $\varphi$. In case of a negative answer, model checking produces a \emph{counterexample}, i.e., a trace $t \in (2^\AP)^\omega$ that is a computation of $\mealy$ that does not satisfy $\varphi$.

To model check a given Mealy machine, we translate the \emph{negation} of the specification into an equivalent automaton, and then check the intersection of the Mealy machine with that automaton for language emptiness. LTL specifications can be represented as B\"uchi automata.

A \emph{nondeterministic B\"uchi automaton} over the alphabet $\Sigma$
is a tuple $\aut=(Q,q_0,\Delta,F)$, where $Q$ is a finite set of
states, $q_0 \in Q$ is an initial state,
$\Delta \subseteq Q \times \Sigma \times Q$ a set of transitions, and
$F \subseteq Q$ a subset of accepting states.  A
nondeterministic B\"uchi automaton accepts an infinite word
$w = w_0w_1w_2\ldots \in \Sigma^\omega$ iff there exists a run $r$ of
$\aut$ on $w$, i.e., an infinite sequence
$r_0r_1r_2\ldots \in Q^\omega$ of states such that $r_0 = q_0$ and
$(r_i, w_i, r_{i+1}) \in \Delta$ for all $i \in \mathbb N$, such that
$r_j \in F$ for infinitely many $j \in \mathbb N$. The set of
sequences accepted by $\aut$ is called the \emph{language}
$\lang(\aut)$ of $\aut$.

\begin{figure}
  \begin{tikzpicture}[auto,semithick]
    \node[state, initial]   (A)                          {$ q_{0} $};
    \node[state, accepting] (B) [below of=A,xshift=-8em] {$ q_{1} $};
    \node[state, accepting] (C) [below of=A] {$ q_{2} $};
    \node[state, accepting] (D) [below of=A,xshift=8em] {$ q_{3} $};

    \path[->]
    (A) edge[loop above] node[above] {$ * $} (A)
    (A) edge node[above left] {$ \request_{1}\overline{\grant_{1}} $} (B)
    (A) edge node[left] {$ \grant_{1}\grant_{2}  $} (C)
    (A) edge node[above right] {$ \request_{2}\overline{\grant_{2}}  $} (D)
    (B) edge[loop below] node[below] {$ \overline{\grant_{1}} $} (B)
    (C) edge[loop below] node[below] {$ * $} (C)
    (D) edge[loop below] node[below] {$ \overline{\grant_{2}} $} (D)
    ;
  \end{tikzpicture}
\caption{%
Nondeterministic B\"uchi automaton corresponding to the negation of the arbiter specification. The states depicted as double circles ($q_{1}$, $q_{2}$, and $q_{3}$) are the accepting states in $F$. The abbreviations $\request_1\overline{\grant_1}$,  $\grant_1\grant_2$, $\request_2\overline{\grant_2}$, $\overline{\grant_1}$, $\overline{\grant_2}$ are used to indicate, in Boolean notation, letters of the alphabet $2^\AP$. E.g., $\request_1\overline{\grant_1}$ represents the letters $ \set{ \request_1, \request_2, \grant_2}$, $ \set{ \request_{1}, \request_{2} }$, $ \set{ \request_{1}, \grant_{2} }$, and $ \set{ \request_{1} } $. The symbol $*$ represents all letters of the alphabet, i.e., all  subsets of $\{\request_1, \request_2, \grant_1, \grant_2 \}$.}
\label{fig:nondetbuchi}
\end{figure}
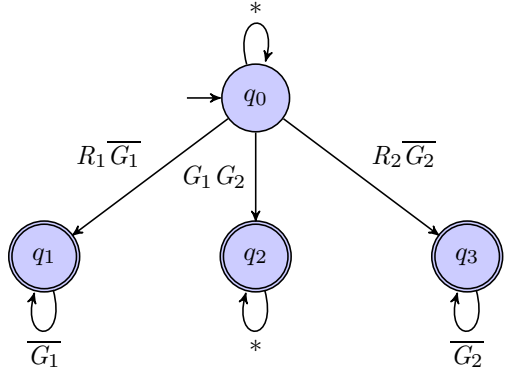

\begin{example}
  Consider the negation of the arbiter specification from Example~\ref{example:arbiter}, i.e., the LTL formula
\[
\begin{array}{ll}
& \LTLdiamond\, (\grant_1 \wedge \grant_2)\\
\vee & \LTLdiamond\, (\request_1 \wedge \LTLsquare \neg \grant_1)\\
\vee & \LTLdiamond\, (\request_2 \wedge \LTLsquare \neg \grant_2)\ .\\
\end{array}
 \]
A nondeterminstic B\"uchi automaton that accepts exactly the traces that satisfy this formula, i.e., all traces that violate the arbiter specification, is shown in Fig.~\ref{fig:nondetbuchi}.
\end{example}

Let $\aut_{\neg
  \varphi}=(Q_{\neg \varphi},q^0_{\neg \varphi},\Delta_{\neg
  \varphi},F_{\neg \varphi})$ be a B\"uchi automaton that accepts all
sequences in $(2^\AP)^\omega$ that satisfy $\neg \varphi$, and therefore
violate $\varphi$.

In model checking, we verify the Mealy machine~$\mealy$ against a specification $\varphi$ by building the product~$\mealy \produ \aut_{\neg \phi}$ of the Mealy machine~$\mealy=(S,s_0,\delta,\gamma)$ over inputs $I$ and outputs $O$, and the B\"uchi automaton $\aut_{\neg \phi}=(Q_{\neg \varphi},q^{0}_{\neg \varphi},\Delta_{\neg
  \varphi},F_{\neg \varphi})$ with alphabet $2^{I\cup O}$.
The product is a directed graph $(V,E)$ with vertices
$V = T \times Q$  and edges
$E \subseteq V \times V$, where $(\tuple{s,q},\tuple{s',q'}) \in E$ iff there is an input $\vec{i} \in 2^I$ such that $\delta(s,\vec{i}) = s' \text{ and } q' \in \Delta_{\neg\varphi}(q, \vec{i} \cup \gamma(s,\vec{i}))$.
The Mealy machine satisfies $\varphi$ iff there is no path in  $\mealy \produ \aut_{\neg \phi}$ that visits an accepting state of $ \aut_{\neg \phi}$ infinitely often.

\begin{figure}
  \begin{center}
    \begin{tikzpicture}[->,>=stealth',shorten >=1pt,auto,thick,scale=1,transform shape]
      \tikzstyle{state}=[rounded corners,rectangle, draw]

      \node[state,initial below,fill=blue!20] (t0q0) {$ \tuple{t_0,q_0}$};
      \node[state,fill=blue!20] (t1q0) [right of=t0q0] {$ \tuple{t_1,q_0}$};
      \node[state,fill=blue!20] (t1q3) [left of=t0q0] {$ \tuple{t_1,q_3}$};
      \node[state,fill=blue!20] (t0q1) [right of=t1q0] {$ \tuple{t_0,q_1}$};

      \path[->]
      (t0q0) edge[loop above] (t0q0)
      (t0q0) edge (t1q3)
      (t0q0) edge[bend left=10] (t1q0)
      (t1q0) edge[loop above] (t1q0)
      (t1q0) edge (t0q1)
      (t1q0) edge[bend left=10] (t0q0)
      ;
    \end{tikzpicture}
  \end{center}
  \vspace{-2em}
\caption{%
  Product of the simple Mealy machine shown on the right in Fig.~\ref{fig:mealymachines} with the B\"uchi automaton from Fig.~\ref{fig:nondetbuchi}.}
\label{fig:crossproduct}
\end{figure}
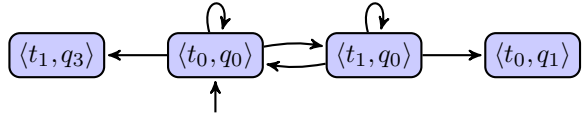

\begin{example}
  Figure~\ref{fig:crossproduct} shows the product $\mealy \produ \aut_{\neg \varphi}$ of the small Mealy machine~$\mealy$ shown on the right in Fig.~\ref{fig:mealymachines} with the B\"uchi automaton $\aut_{\neg \varphi}$ from Fig.~\ref{fig:nondetbuchi}. The only infinite paths are the self-loops from $\tuple{t_0,q_0}$ and $\tuple{t_1,q_0}$ and the path that oscillates forever between $\tuple{t_0,q_0}$ and $\tuple{t_1,q_0}$. These paths do not visit any accepting states. $\mealy$ thus satisfies~$\varphi$.
\end{example}

\section{Game-based Synthesis}
\label{sec:games}

In the classic game-based approach to synthesis~\cite{Rabin/72/Automata}, the problem is analyzed in terms of a two-player game.
The game is played between two players: the input player \emph{Player~$ I $} determines the inputs to the system with the goal of \emph{violating} the specification. The output player \emph{Player~$ O $} controls the outputs of the system with the goal of \emph{satisfying} the specification.
A winning strategy for Player~$I$ can be translated into an implementation that is guaranteed to satisfy the specification. To solve the synthesis problem, we must therefore check whether Player~$I$ has a winning strategy.

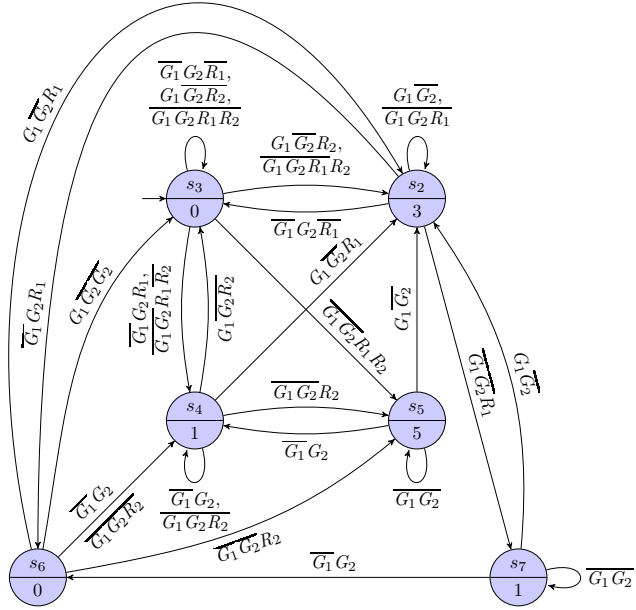
\begin{figure}
\centering
\scalebox{0.7}{
\begin{tikzpicture}[node distance=12em]
  \clip (-4,-8) rectangle (8.4,4);

  \node[state,initial,circle split] (A) {$ s_{3} $ \nodepart{lower} 0};
  \node[state,circle split] (B) [right of=A] {$ s_{2} $ \nodepart{lower} 3};
  \node[state,circle split] (C) [below of=A] {$ s_{4} $ \nodepart{lower} 1};
  \node[state,circle split] (D) [below of=B] {$ s_{5} $ \nodepart{lower} 5};
  \node[state,circle split] (E) [below right of=D,xshift=-3em] {$ s_{7} $ \nodepart{lower} 1};
  \node[state,circle split] (F) [below left of=C] {$ s_{6} $ \nodepart{lower} 0};
  
  \path[->]
  (A) edge[loop above] node[above] {
    \begin{tabular}{c}
      $ \overline{\grant_{1}}\grant_{2}\overline{\request_{1}}, $ \\
      $ \grant_{1}\overline{\grant_{2}}\overline{\request_{2}}, $ \\
      $ \overline{\grant_{1}}\overline{\grant_{2}}\overline{\request_{1}}\overline{\request_{2}} $
    \end{tabular}
    } (A)
  (A) edge[bend right=10] node[above, rotate=90] {
    \begin{tabular}{c}
      $ \overline{\grant_{1}}\grant_{2}\request_{1} $, \\
      $ \overline{\grant_{1}}\overline{\grant_{2}}\request_{1}\overline{\request_{2}} $
    \end{tabular}
    } (C)
  (A) edge[bend left=10] node[above] {
    \begin{tabular}{c}
      $ \grant_{1}\overline{\grant_{2}}\request_{2} $, \\
      $ \overline{\grant_{1}}\overline{\grant_{2}}\overline{\request_{1}}\request_{2} $
    \end{tabular}
    } (B)
  (A) edge node[above, rotate=-45,xshift=3em] {
    $  \overline{\grant_{1}}\overline{\grant_{2}}\request_{1}\request_{2} $
    } (D)
  (B) edge[loop above] node[above] {
    \begin{tabular}{c}
      $ \grant_{1}\overline{\grant_{2}}, $ \\
      $ \overline{\grant_{1}}\overline{\grant_{2}}\overline{\request_{1}} $
    \end{tabular}
    } (B)   
  (B) edge[bend left=10] node[below] {
    $ \overline{\grant_{1}}\grant_{2}\overline{\request_{1}} $
    } (A)
  (B) edge node[above,rotate=-75] {
    $ \overline{\grant_{1}}\overline{\grant_{2}}\request_{1} $
    } (E)
  (B) edge[out=130,in=90,min distance=23em] node[pos=0.8,above,rotate=85] {
    $ \overline{\grant_{1}}\grant_{2}\request_{1} $
    } (F)
  (C) edge[loop below] node[below] {
    \begin{tabular}{c}
      $ \overline{\grant_{1}}\grant_{2}, $ \\
      $ \overline{\grant_{1}}\overline{\grant_{2}}\overline{\request_{2}} $
    \end{tabular}
    } (C)   
  (C) edge[bend right=10] node[below,rotate=90] {
    $ \grant_{1}\overline{\grant_{2}}\overline{\request_{2}} $
    } (A) 
  (C) edge node[above,rotate=45,xshift=3em] {
    $ \grant_{1}\overline{\grant_{2}}\request_{1} $
    } (B)
  (C) edge[bend left=10] node[above] {
    $ \overline{\grant_{1}}\overline{\grant_{2}}\request_{2} $
    } (D)
  (D) edge[loop below] node[below] {
    $ \overline{\grant_{1}}\overline{\grant_{2}} $
    } (D)
  (D) edge node[above,rotate=90] {
    $ \grant_{1}\overline{\grant_{2}} $
    } (B)
  (D) edge[bend left=10] node[below] {
    $ \overline{\grant_{1}}\grant_{2} $
    } (C)
  (E) edge[loop right] node[right] {
    $ \overline{\grant_{1}}\overline{\grant_{2}} $
    } (E)
  (E) edge[bend right=20] node[above,rotate=-75] {
    $ \grant_{1}\overline{\grant_{2}} $
    } (B)
  (E) edge node[above,xshift=3em] {
    $ \overline{\grant_{1}}\grant_{2} $
    } (F)
  (F) edge[out=10,in=220] node[below,rotate=20] {
    $ \overline{\grant_{1}} \overline{\grant_{2}}\request_{2} $
    } (D)
  (F) edge node[above,rotate=45,xshift=-1em] {
    $ \overline{\grant_{1}}\grant_{2} $
    } node[below,rotate=45,xshift=-1em] {
    $ \overline{\grant_{1}}\overline{\grant_{2}}\overline{\request_{2}} $
    } (C)
  (F) edge[out=80,in=220] node[above,pos=0.7,rotate=60] {
    $ \grant_{1}\overline{\grant_{2}}\overline{\grant_{2}} $
    } (A)
  (F) edge[out=103,in=120,min distance=24em] node[above,pos=0.4,rotate=64] {
    $ \grant_{1}\overline{\grant_{2}}\request_{1} $
    } (B)
  ;
\end{tikzpicture}
}
\caption{Deterministic parity automaton corresponding to the arbiter
  specification. The colors of the states are shown in the lower
  part of the state labels.}
\label{fig:detparity}
\end{figure}

In order to turn the specification into a game, we translate
the LTL formula into a deterministic automaton that accepts all
traces that satisfy the formula.  An automaton is
\emph{deterministic} if each state and input has unique successor
state, i.e., the set of transitions $ \Delta $ is a total function
from $ Q \times \Sigma $ to $ Q $.
Since
deterministic Büchi automata are not expressive enough to represent
every possible LTL specification, we must use a more expressive acceptance condition
such as the 
parity condition. Whereas a Büchi acceptance
condition identifies a set $F \subseteq S$ of accepting
states, which have to be visited infinitely often, a \emph{parity condition} $c: S \rightarrow \mathbb N$ labels every state with a natural
number. We call such a number the \emph{color} of the state. A run of a
parity automaton is accepting if the smallest color that appears
infinitely often as a label of the states of the run is even. This
introduces a hierarchy in the acceptance condition, as from some
point on, every odd color has to be answered by a smaller even color.
The Büchi acceptance condition is a special case of the parity condition, where the accepting states are colored with 0
and the remaining states are colored with 1.


\begin{example}
  Figure~\ref{fig:detparity} shows a deterministic parity automaton,
  whose language consists of all traces that satisfy the arbiter specification from
  Example~\ref{example:arbiter}. The colors of the states are shown
  in the lower part of the state labels.
\end{example}

The deterministic automaton is then translated into an infinite game over a finite graph.
A \emph{game graph} is a directed graph $ (V,E) $ with vertices $V$ and edges $E$.
The vertices $V = V_{I} \dot\cup V_{O} $ are partitioned into the vertices $V_I$ controlled by Player~$I$ and the
vertices $V_O$ controlled by Player~$O$. A \emph{parity game} $(V,E,c)$ consists of a game graph $(V,E)$ and a
parity condition $c: V \rightarrow \mathbb N$. To play the
game, a token is placed on some initial vertex~$ v $,
which is then moved by the player owning the vertex to one of its
successors~$ v' $, i.e., such that $ (v,v') \in E $. This is repeated ad infinitum, resulting in an infinite sequence of
vertices, called a \emph{play} of the game. If the underlying color
sequence, i.e., the sequence resulting by the reduction of the
vertices to their labels, satisfies the parity condition, Player~$ O $
wins the game, otherwise Player~$ I $ wins the game. 

The game for the synthesis problem is obtained from the deterministic automaton 
by separating the moves of Player~$ I $, namely the choice of the
inputs~$ I $ to the system, and the moves of Player~$ O $, i.e., the choice of the
outputs~$ O $.

We are interested in finding a winning strategy for Player~$O$, i.e., an appropriate
choice of output after every possible prefix of a play. We call such a
prefix a \emph{history} of the play. A
useful property of parity games is that they are \emph{memoryless determined},
which means that if one of the players has a winning strategy, then there
also exists a winning strategy that only depends on the last vertex of
the history, ignoring the previously visited vertices.
For parity games, it is possible to automatically compute the set of
vertices from which Player~$O$ has a winning strategy. This set of
vertices is called the \emph{winning region}.
If the vertex corresponding to the initial state of the automaton is in the
winning region, then there exists a solution to the synthesis problem.

\begin{figure}
\centering
\begin{tikzpicture}[scale=0.9]
  \node[win] at (3.5,4.2) {}; 
  \node[win] at (-2.9,0.2) {}; 
  \node[win] at (5.3,2.8) {};
  \node[win] at (5.6,5.1) {};
  \node[win] at (-0.9,5.5) {};
  \node[win] at (-2.3,3.4) {}; 
  \node[win] at (-5.6,0.9) {}; 
  \node[win] at (0.8,4.2) {};
  \node[win] at (-0.25,-3) {}; 
  \node[win] at (-5.6,-1.6) {};
  \node[win] at (-4.1,-1.6) {}; 
  \node[win] at (5,0.25) {};
  \node[win] at (3.2,-5.35) {};
  \node[win] at (-1.1,-5.5) {};  
  \node[win] at (5,-5.4) {};  
  \node[win] at (4.95,-3.55) {};
  \node[win] at (-2.5,-3.8) {}; 
  \node[win] at (1.8,-2.8) {};

  \node[pl1,scale=0.7,initial above] (S0) at (3.5,4.2) {0}; 
  \node[pl0,scale=0.7] (S1) at (-2.9,0.2) {6}; 
  \node[pl0,scale=0.7] (S2) at (5.3,2.8) {6};
  \node[pl0,scale=0.7] (S3) at (5.6,5.1) {2};
  \node[pl0,scale=0.7] (S4) at (-0.9,5.5) {4};
  \node[pl1,scale=0.7] (S5) at (-2.3,3.4) {1}; 
  \node[pl1,scale=0.7] (S6) at (0.8,0.9) {5}; 
  \node[pl0,scale=0.7] (S7) at (-1.1,2.3) {8}; 
  \node[pl0,scale=0.7] (S8) at (-5.6,0.9) {3}; 
  \node[pl0,scale=0.7] (S9) at (0.8,4.2) {7};
  \node[pl1,scale=0.7] (S10) at (-0.25,-3) {1}; 
  \node[pl1,scale=0.7] (S11) at (-5.6,-1.6) {7};
  \node[pl0,scale=0.7] (S12) at (-4.1,-1.6) {9}; 
  \node[pl0,scale=0.7] (S13) at (5,0.25) {3};
  \node[pl0,scale=0.7] (S14) at (3.2,-5.35) {1};
  \node[pl1,scale=0.7] (S15) at (-1.1,-5.5) {8};  
  \node[pl1,scale=0.7] (S16) at (5,-5.4) {3};  
  \node[pl0,scale=0.7] (S17) at (4.95,-3.55) {1};
  \node[pl0,scale=0.7] (S18) at (-2.5,-3.8) {0}; 
  \node[pl0,scale=0.7] (S19) at (1.8,-2.8) {4};

  \path[->]
  (S0) edge[bend left=10] (S4) edge[bend left=10] (S3) edge[bend left=10] (S2) edge (S1)
  (S1) edge[strat] (S5) edge (S6) edge (S11) edge (S10)
  (S2) edge[bend left=10,strat] (S0) edge (S6) edge (S10)
  (S3) edge[bend left=10,strat] (S0) edge (S6)
  (S4) edge (S5) edge[bend left=10,strat] (S0) edge (S6)
  (S5) edge[bend left=10] (S8) edge[bend left=10] (S9)
  (S6) edge[bend left=10] (S7)
  (S7) edge[bend left=10] (S6)
  (S8) edge[bend left=10] (S5) edge (S6) edge (S11) edge[strat] (S10)
  (S9) edge[bend left=10,strat] (S5) edge (S6) edge (S0)
  (S10) edge[bend left=10] (S13) edge[bend left=10] (S14)
  (S11) edge[bend left=10] (S12)
  (S12) edge[bend left=10] (S11) edge (S5) edge (S6) edge[strat] (S10)
  (S13) edge[bend left=10,strat] (S10) edge (S6) edge (S0)
  (S14) edge[bend left=10] (S10) edge (S6) edge[strat] (S15) edge (S16)
  (S15) edge (S18) edge (S19)
  (S16) edge[bend left=10] (S17)
  (S17) edge[bend left=10] (S16) edge (S6) edge (S10) edge[strat] (S15)
  (S18) edge (S11) edge (S6) edge (S10) edge (S15) edge[strat] (S5)
  (S19) edge (S6) edge[strat] (S0) edge (S5)
  ;
\end{tikzpicture}
\caption{Parity game resulting from the deterministic parity automaton depicted in Fig.~\ref{fig:detparity}. Vertices controlled by Player~$ I $ are
  depicted as rectangles, vertices controlled by Player~$ O $ as
  circles.  The highlighted states mark the winning region of Player~$ O $. The winning strategy is indicated by the thick edges.}
\label{fig:game}
\end{figure}
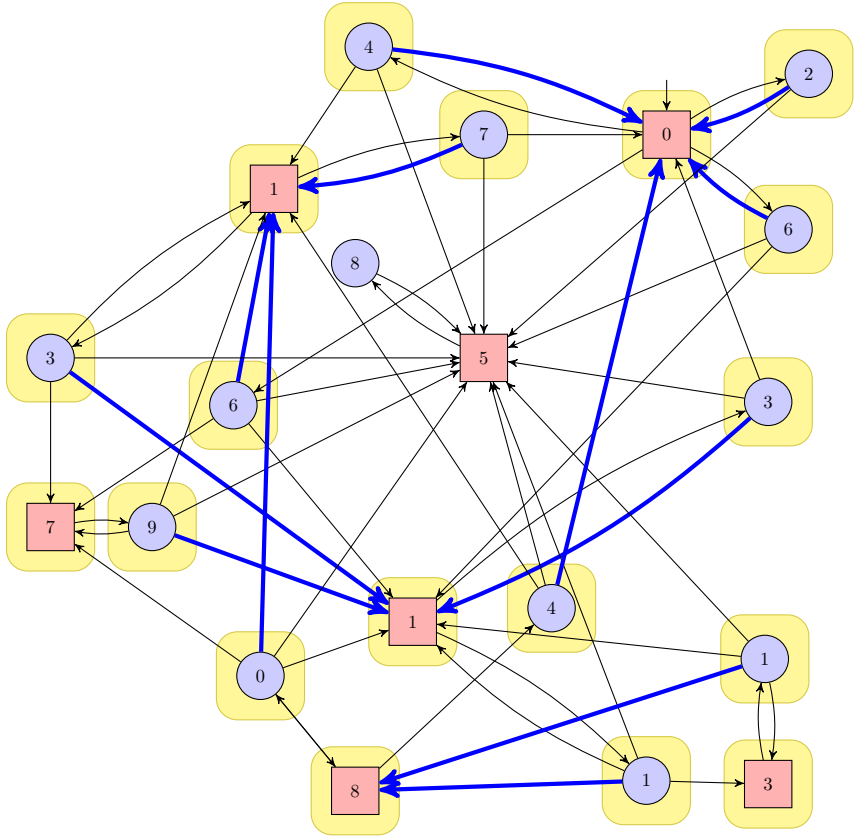

\begin{example}
  Figure~\ref{fig:game} shows the parity game for the synthesis problem
  of the arbiter specification from Example~\ref{example:arbiter}.
  The game was constructed by first translating the LTL formula into
  the deterministic automaton shown in
  Fig.~\ref{fig:detparity}, and then separating the moves of the
  input and output players.
  In Fig.~\ref{fig:game}, vertices controlled by Player~$ I $ are
  depicted as rectangles, vertices controlled by Player~$ O $ as
  circles. The winning region of Player~$ O $ is marked by the highlighting. 
   Since the initial vertex
  is in the winning region, the specification can be
  realized. The (memoryless) winning strategy
  is indicated by the thick edges.
\end{example}

Game-based synthesis is asymptotically optimal in the size of the input. However, the synthesized implementations are often much larger than necessary. Compare, for example, the size of the winning strategy in Fig.~\ref{fig:game} with the small Mealy machine on the right in Fig.~\ref{fig:mealymachines}.

\section{Bounded Synthesis}
\label{sec:boundedsynthesis}

In bounded synthesis~\cite{boundedjournal}, we set a bound on the number of states of the synthesized Mealy machine. By incrementally increasing the bound, we can use bounded synthesis to find a Mealy machine with a minimal number of states. The Mealy machine is found as a solution of a constraint system.
To ensure that all solutions of the constrain system satisfy the specification, we encode not only the states, transitions, and outputs of the Mealy machine, but, additionally, an annotation of the states of the Mealy machine that ensures that the given LTL specification is satisfied.
This annotation essentially ensures that the model checking of the Mealy machine succeeds, i.e., that the language of the product with the B\"uchi automaton corresponding to the negation of the specification is empty.

Let $\tuple{V,E}$ be the product of a Mealy machine $\mealy$ and a B\"uchi automaton $\aut_{\neg \varphi}$ for the negation of the specification.
An \emph{annotation} $\lambda : S \times Q \rightarrow \{\bot\} \cup \mathbb N$ is a function that maps nodes from the run graph to either unreachable $\bot$ or a natural number $k$.
An annotation is valid if it satisfies the following conditions:
\begin{itemize}
  \item the initial vertex $\tuple{s_0,q_0}$ is labeled by a natural number: $\lambda(t_0,q_0) \neq \bot$, and
  \item if a vertex $\tuple{s,q}$ is annotated with a natural number, i.e., $\lambda(t,q) = k \neq \bot$, then for every $\vec{i} \in 2^I$ and $q' \in \Delta_{\neg_\varphi}(q, \vec{i} \cup \gamma(s,\vec{i}),q')$, the successor pair $\tuple{\tau(s,\vec{i}),q')}$ is annotated with a \emph{greater or equal} number, which needs to be \emph{strictly} greater if $q'$ is a rejecting state. That is, $\lambda(t',q') > k$ if $q' \in F$ and $\geq$$\lambda(t',q') \geq k$ otherwise.
\end{itemize}

\begin{figure}[t]
  \begin{center}
    \begin{tikzpicture}[->,>=stealth',shorten >=1pt,auto,thick,scale=1,transform shape]
      \tikzstyle{twostate}=[rounded corners,rectangle split, rectangle split parts=2, draw]

      \node[twostate,initial below,fill=blue!20] (t0q0) {$ \tuple{t_0,q_0}$ \nodepart{two} $\lambda: 0$};
      \node[twostate,fill=blue!20] (t1q0) [right of=t0q0] {$ \tuple{t_1,q_0}$ \nodepart{two} $\lambda: 0$};
      \node[twostate,fill=blue!20] (t1q3) [left of=t0q0] {$ \tuple{t_1,q_3}$ \nodepart{two} $\lambda: 1$};
      \node[twostate,fill=blue!20] (t0q1) [right of=t1q0] {$ \tuple{t_0,q_1}$ \nodepart{two} $\lambda: 1$};
      \node[twostate,fill=blue!20] (t0q3) [above of=t1q3] {$ \tuple{t_0,q_3}$ \nodepart{two} $\lambda: \bot$};
      \node[twostate,fill=blue!20] (t0q2) [above of=t0q0] {$ \tuple{t_0,q_2}$ \nodepart{two} $\lambda: \bot$};
      \node[twostate,fill=blue!20] (t1q2) [above of=t1q0] {$ \tuple{t_1,q_2}$ \nodepart{two} $\lambda: \bot$};
      \node[twostate,fill=blue!20] (t1q1) [above of=t0q1] {$ \tuple{t_1,q_1}$ \nodepart{two} $\lambda: \bot$};

      \path[->]
      (t0q0) edge[loop above] (t0q0)
      (t0q0) edge (t1q3)
      (t0q0) edge[bend left=10] (t1q0)
      (t1q0) edge[loop above] (t1q0)
      (t1q0) edge (t0q1)
      (t1q0) edge[bend left=10] (t0q0)
      ;
    \end{tikzpicture}
  \end{center}
  \vspace{-2em}
  \caption{Annotated product of the simple Mealy machine shown on the right in Fig.~\ref{fig:mealymachines} with the B\"uchi automaton from Fig.~\ref{fig:nondetbuchi}.}
  \label{fig:run-graph-arbiter}
\end{figure}
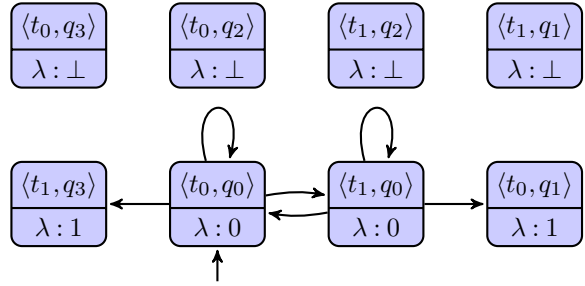

\begin{example}
  Figure~\ref{fig:run-graph-arbiter} shows the
annotated product of the simple Mealy machine from the right in Fig.~\ref{fig:mealymachines} with the B\"uchi automaton from Fig.~\ref{fig:nondetbuchi}.
  One can verify that the annotation is correct by checking every edge individually.
  For example, the annotation has to increase from $\tuple{t_0,q_0} \rightarrow \tuple{t_1,q_3}$ and from $\tuple{t_1,q_0} \rightarrow \tuple{t_0,q_1}$ as $q_1$ and $ q_{3} $ are rejecting.
\end{example}



The existence of a Mealy machine with a corresponding annotation of the product graph can be expressed as a propositional constraint. For this purpose, we encode the Mealy machine and the annotation with Boolean variables.

\begin{itemize}

\item $ \ttrans{t}{\nu}{t'} $ for all $ t,t' \in S $ and
  $ \nu \in 2^{\inputs} $, for the transition function $\delta\colon S \times 2^I \rightarrow S$ of the Mealy machine $ \mealy = (S, s_{0}, \delta, \gamma) $.

\item $ \tlabel{t}{\nu}{x} $ for all $ t \in S $,
  $ \nu \in 2^{\inputs} $ and $ x \in \outputs $, for the output function $\delta\colon S \times 2^I \rightarrow S$.

\item $ \rgstate{t}{q} $ for all $ t \in T $ and $ q \in Q $, to
  encode the reachable states of the product graph~$ G $ of $ \mealy $ and
  $ \aut_{\neg \varphi} $, i.e., those state pairs $\langle t,q \rangle$ where $\lambda(t,q) \neq \bot$.

\item $ \annotationd{t}{q}{i} $ for all $ t \in T $, $ q \in Q $ and
  $ 0 < i \leq \log(n \cdot k) $, where $n$ is the bound on the size of the Mealy machine and $k$ is the number of states of the B\"uchi automaton. The variables encode the numerical annotation of a state pair
  $ (t,q) $ of $ G $. We use a logarithmic number of bits to
  encode the annotated value in binary. 
\end{itemize}

\noindent Given an LTL formula~$ \varphi $ and a bound
$ n $ on the states of the Mealy machine, we solve the
bounded synthesis problem by checking the satisfiability of the propositional
formula~$ \curlyF_{BS}(\varphi,n) $, consisting of the following
constraints:

\begin{itemize}

  %
%

\item The pair of initial states~$ \langle s_{0},q_{0}\rangle $ for some
  arbitrary, but fixed, $ s_{0} $ is reachable and annotated with 1. 
  \begin{equation*}
    \textsc{rgstate}(s_0,q_0) \wedge \annotation{1}{1} = 1 
  \end{equation*}

\goodbreak

\item Each annotation of a vertex of the product graph bounds the number
  of visited accepting states, not counting the current vertex itself:
  \begin{equation*}
    \bigwedge\limits_{t \in T,\, q \in Q} \rgstate{t}{q} \rightarrow 
    \bigwedge\limits_{\sigma \in 2^{\Sigma}} \textit{output}(t,\sigma)
    \rightarrow \bigwedge\limits_{t' \in T} \ttrans{t}{\inputs \cap 
      \sigma}{t'} \rightarrow 
  \end{equation*}
  \begin{equation*}
    \qquad \bigwedge\limits_{q' \in \Delta(q,\sigma)} \rgstate{t'}{q'} 
    \wedge \annotation{t}{q} \prec_{q}  \annotation{t'}{q'}
  \end{equation*}
  where $ \prec_{q} $ equals $ < $ if $ q \in R $ and equals $ \leq $
  otherwise. The
  formula~$ \textit{output}(t,\sigma) $ ensures that the output corresponds to the output function of the Mealy machine, i.e.,
  \[ \textit{output}(t,\sigma) = \bigwedge_{x \in \outputs \cap
    \sigma}\, \tlabel{t}{\inputs \cap \sigma}{x} \wedge \bigwedge_{x
    \in \outputs \smallsetminus \sigma} \neg \tlabel{t}{\inputs \cap
    \sigma}{x} .\]

  \medskip

\end{itemize}

\begin{theorem}[Bounded Synthesis~\cite{boundedjournal}]
  For an LTL formula $\varphi$ and a bound $ n \in \nats $, the propositional formula~$ \curlyF_{BS}(\varphi,n) $ is
  satisfiable if and only if there is a Mealy machine~$ \mealy $ with
  $ \size{\mealy} = n $ that satisfies $ \varphi $.
\end{theorem}

The propositional constraint can be solved by a standard SAT
solver. In addition to the encoding as a propositional constraint, the
bounded synthesis problem has also been reduced to the satisfiability of quantified Boolean
formulas (QBF) and dependency quantified Boolean formulas
(DQBF)~\cite{Faymonville2017}, as well as to satisfability modulo theories (SMT)~\cite{Finkbeiner+Schewe/07/smt}. Such encodings are more concise than
the encoding as a Boolean formula. Even though the satisfiability
problems of these logics are more expensive than propositional
satisfiability, in particular the QBF encoding has proven advantageous
in experiments (cf.~\cite{Faymonville2017b}).

Another powerful optimization is \emph{lazy synthesis}~\cite{Finkbeiner+Jacobs/12/Lazy}, which avoids the
full construction of the constraint system.  Lazy synthesis alternates
between constraint solving, where a model is constructed for an
incomplete constraint system, and verification, where errors in the
previously constructed model are identified and used to extend the
constraint system.

\section{Bounded Cycle Synthesis}
\label{sec:boundedcyclesynthesis}

Bounded cycle synthesis~\cite{FK:2016} extends bounded synthesis by bounding not
only the number of states, but also the number of cycles of the Mealy machine. Bounded cycle synthesis allows us to find
implementations that are not only small but also structurally simple.
%
%
%
%
%
A cycle is a path of a Mealy machine that ends in the same state it started in.
Even Mealy machines with a small number of states can have many cycles: the number of cycles can
be exponential in the number of states. The explosion of the number of circles is in fact worse than the explosion of
the number of states: while a realizable LTL formula has an implementation with at most doubly exponentially many states,
there exist LTL formulas where the number of cycles in the Mealy machine is triply exponential~\cite{FK:2016}. 
This makes the number of cycles a particularly interesting metric for output-sensitive synthesis algorithms.




Let $ G = (V,E) $ be a directed graph. A \textit{(simple) cycle}~$ c $
of $ G $ is a a tuple~$ (C,\eta) $, consisting of a non-empty set~$ C
\subseteq V $ and a bijection $ \eta \colon C \mapsto C $ such that 

\begin{itemize}

\item $ \forall v \in C.\ (v,\eta(v)) \in E $ and 

\item $ \forall v \in C.\ n \in \nats.\ \eta^{n}(v) = v
  \ \Leftrightarrow \ n \!\!\mod \size{C} = 0 $,

\end{itemize}

\noindent where $ \eta^{n} $ denotes $ n $ times the application of
$ \eta $.  In other words, a cycle of $ G $ is a path through $ G $
that starts and ends at the same vertex and visits every vertex of
$ V $ at most once. We say that a cycle~$ c = (C,\eta) $ has length
$ n $ iff $ \size{C} = n $.

We extend the notion of a cycle of a graph~$ G $ to Mealy
machines~$ \mealy = (T, t_{I}, \delta, \lambda) $, such that
$ c $ is a cycle of $ \mealy $ iff $ c $ is a cycle of the graph
$ (T,E) $ for
$ E = \set{ (t,t') \mid \exists \nu \in 2^{\inputs}.\ \delta(t,\nu) =
  t' } $.
Thus, we ignore the input labels of the edges of $ \mealy $.  The set
of all cycles of a Mealy machine~$ \mealy $ is denoted by
$ \mathcal{C}(\mealy) $.

\subsection{Counting Cycles}

A classical algorithm for counting the number of cycles of a directed
graph is due to Tiernan~\cite{Tiernan:1970}. We review this
algorithm here as a preparation for the bounded cycle synthesis
encoding.


\goodbreak

\smallskip

\paragraph{Algorithm 1.} Given a directed graph~$ G = (V,E) $, we
count the cycles of $ G $ using the following algorithm:

\begin{enumerate}

\item[(1)] Initialize the cycle counter~$ c $ to $ c := 0 $ and some
  set~$ P $ to $ P := \emptyset $.

\item[(2)] Pick some arbitrary vertex~$ v_{r} $ of $ G $, set
  $ v := v_{r} $ and $ P := \set{ v_{r} } $.

\item[(3)] For all edges $ (v,v') \in E $, with
  $ v' \notin P \setminus \set{ v_{r} } $:

\begin{enumerate}

\item[(3a)] If $ v' = v_{r} $, increase $ c $ by one.

\item[(3b)] Oherwise, set $ v := v' $, add $ v' $ to $ P $ and recursively execute
  (3). Afterwards, reset $ P $ to its value before the recursive call.

\end{enumerate}

\item[(4)] Obtain the sub-graph~$ G' $, by removing $ v_{r} $ from $ G $:

\begin{enumerate}

\item[(4a)] If $ G' $ is empty, return $ c $.

\item[(4b)] Otherwise, continue from (2) with $ G' $.

\end{enumerate}

\end{enumerate}

\noindent The algorithm starts by counting all cycles that contain
the first picked vertex~$ v_{r} $. This is done by an unfolding of the
graph into a tree, rooted in $ v_{r} $, such that there is no repetition
of a vertex on any path from the root to a leaf. The number of
vertices that are connected to the root by an edge of $ E $ then
represents the corresponding number of cycles through $ v_{r} $. The
remaining cycles of $ G $ do not contain $ v_{r} $ and, thus, are
cycles of the sub-graph $ G' $ without $ v_{r} $, as well. Hence, we
count the remaining cycles by recursively counting the cycles of
$ G' $. The algorithm terminates as soon as $ G' $ becomes empty.

The algorithm, as described so far, has the disadvantage that the number of 
unfolded trees is exponential in the size of the
graph, even if none of their vertices is connected to the root, i.e.,
even if there is no cycle to be counted. This drawback can be
avoided by first reducing the graph to all its strongly connected
components (SCCs) and then counting the cycles of each SCC
separately~\cite{Weinblatt:1972,Johnson:1975}. This reduction is
sound, as a cycle never leaves an SCC of the graph.

The improved algorithm is exponential in
the size of $ G $, and linear in the number of cycles~$ m $.
Furthermore, the time between two detections of a cycle, during the
execution, is bounded linear in the size of~$ G $.

\begin{figure}
  \begin{tikzpicture}
    \node[draw, inner sep=3pt,rounded corners=3] (S1) at (0,0) {
      \begin{tikzpicture}
        \node at (0,-1.8) {$ c := 0 $, $ v := s_{0} $, $ v_{r} := s_{0} $};
        \node at (0,-2.2) {$ P := \set{ s_{0} } $};
        \node[anchor=west] at (-1.8,1.4) {\textbf{(1) + (2)}};
        \node at (0,0) {
          \scalebox{0.8} {
            \begin{tikzpicture}[auto,thick]
              \node[state,marked] (A)                          {$ s_{0} $};
              \node[state]          (B) [below of=A,xshift=-4em] {$ s_{1} $};
              \node[state]          (C) [below of=A,xshift=4em]  {$ s_{2} $};

              \path[->]
              (A) edge[loop above] (A)
              (A) edge[out=0,in=75] (C)
              (B) edge (A)
              (B) edge[bend left=10] (C)
              (C) edge[bend left=10] (B)   
              (C) edge (A)
              ;
            \end{tikzpicture}
          }
        };
      \end{tikzpicture}
    };

    \node[draw, inner sep=3pt,rounded corners=3] (S2) at (4,0) {
      \begin{tikzpicture}
        \node at (0,-1.8) {$ c := 1 $, $ v := s_{0} $, $ v_{r} := s_{0} $};
        \node at (0,-2.2) {$ P := \set{ s_{0} } $};

        \node[anchor=west] at (-1.8,1.4) {\textbf{(3) + (3a)}};
        \node[anchor=west] at (-1.8,0.95) {$(s_{0}, s_{0}) $};
        \node at (0,0) {
          \scalebox{0.8} {
            \begin{tikzpicture}[auto,thick]
              \node[state,marked] (A)                          {$ s_{0} $};
              \node[state]          (B) [below of=A,xshift=-4em] {$ s_{1} $};
              \node[state]          (C) [below of=A,xshift=4em]  {$ s_{2} $};

              \path[->]
              (A) edge[loop above,ultra thick,red] (A)
              (A) edge[out=0,in=75] (C)
              (B) edge (A)
              (B) edge[bend left=10] (C)
              (C) edge[bend left=10] (B)   
              (C) edge (A)
              ;
            \end{tikzpicture}
          }
        };
      \end{tikzpicture}
    };

    \node[draw, inner sep=3pt,rounded corners=3] (S3) at (8,0) {
      \begin{tikzpicture}
        \node at (0,-1.8) {$ c := 1 $, $ v := s_{2} $, $ v_{r} := s_{0} $};
        \node at (0,-2.2) {$ P := \set{ s_{0}, s_{2} } $};

        \node[anchor=west] at (-1.8,1.4) {\textbf{(3) + (3b)}};
        \node[anchor=west] at (-1.8,0.95) {$(s_{0}, s_{2}) $};
        \node at (0,0) {
          \scalebox{0.8} {
            \begin{tikzpicture}[auto,thick]
              \node[state,marked] (A)                          {$ s_{0} $};
              \node[state]          (B) [below of=A,xshift=-4em] {$ s_{1} $};
              \node[state]          (C) [below of=A,xshift=4em]  {$ s_{2} $};

              \path[->]
              (A) edge[loop above, gray!50] (A)
              (A) edge[out=0,in=75,ultra thick,red] (C)
              (B) edge (A)
              (B) edge[bend left=10] (C)
              (C) edge[bend left=10] (B)   
              (C) edge (A)
              ;
            \end{tikzpicture}
          }
        };
      \end{tikzpicture}
    };

    \node[draw, inner sep=3pt,rounded corners=3] (S4) at (0,-4.6) {
      \begin{tikzpicture}
        \node at (0,-1.8) {$ c := 2 $, $ v := s_{2} $, $ v_{r} := s_{0} $};
        \node at (0,-2.2) {$ P := \set{ s_{0},s_{2} } $};

        \node[anchor=west] at (-1.8,1.4) {\textbf{(3) + (3a)}};
        \node[anchor=west] at (-1.8,0.95) {$ (s_{2}, s_{0}) $};
        \node at (0,0) {
          \scalebox{0.8} {
            \begin{tikzpicture}[auto,thick]
              \node[state,marked] (A)                          {$ s_{0} $};
              \node[state]          (B) [below of=A,xshift=-4em] {$ s_{1} $};
              \node[state]          (C) [below of=A,xshift=4em]  {$ s_{2} $};

              \path[->]
              (A) edge[loop above, gray!50] (A)
              (A) edge[out=0,in=75,ultra thick,red] (C)
              (B) edge (A)
              (B) edge[bend left=10] (C)
              (C) edge[bend left=10] (B)   
              (C) edge[ultra thick,red] (A)
              ;
            \end{tikzpicture}
          }
        };
      \end{tikzpicture}
    };

    \node[draw, inner sep=3pt,rounded corners=3] (S5) at (4,-4.6) {
      \begin{tikzpicture}
        \node at (0,-1.8) {$ c := 2 $, $ v := s_{1} $, $ v_{r} := s_{0} $};
        \node at (0,-2.2) {$ P := \set{ s_{0},s_{1},s_{2} } $};

        \node[anchor=west] at (-1.8,1.4) {\textbf{(3) + (3b)}};
        \node[anchor=west] at (-1.8,0.95) {$ (s_{2}, s_{1}) $};
        \node at (0,0) {
          \scalebox{0.8} {
            \begin{tikzpicture}[auto,thick]
              \node[state,marked] (A)                          {$ s_{0} $};
              \node[state]          (B) [below of=A,xshift=-4em] {$ s_{1} $};
              \node[state]          (C) [below of=A,xshift=4em]  {$ s_{2} $};

              \path[->]
              (A) edge[loop above, gray!50] (A)
              (A) edge[out=0,in=75,ultra thick,red] (C)
              (B) edge (A)
              (B) edge[bend left=10] (C)
              (C) edge[bend left=10,ultra thick,red] (B)   
              (C) edge[gray!50] (A)
              ;
            \end{tikzpicture}
          }
        };
      \end{tikzpicture}
    };

    \node[draw, inner sep=3pt,rounded corners=3] (S6) at (8,-4.6) {
      \begin{tikzpicture}
        \node at (0,-1.8) {$ c := 3 $, $ v := s_{1} $, $ v_{r} := s_{0} $};
        \node at (0,-2.2) {$ P := \set{ s_{0},s_{1},s_{2} } $};

        \node[anchor=west] at (-1.8,1.4) {\textbf{(3) + (3a)}};
        \node[anchor=west] at (-1.8,0.95) {$ (s_{1}, s_{0}) $};
        \node at (0,0) {
          \scalebox{0.8} {
            \begin{tikzpicture}[auto,thick]
              \node[state,marked] (A)                          {$ s_{0} $};
              \node[state]          (B) [below of=A,xshift=-4em] {$ s_{1} $};
              \node[state]          (C) [below of=A,xshift=4em]  {$ s_{2} $};

              \path[->]
              (A) edge[loop above, gray!50] (A)
              (A) edge[out=0,in=75,ultra thick,red] (C)
              (B) edge[ultra thick,red] (A)
              (B) edge[bend left=10] (C)
              (C) edge[bend left=10,ultra thick,red] (B)   
              (C) edge[gray!50] (A)
              ;
            \end{tikzpicture}
          }
        };
      \end{tikzpicture}
    };

    \node[draw, inner sep=3pt,rounded corners=3] (S7) at (0,-9.2) {
      \begin{tikzpicture}
        \node at (0,-1.8) {$ c := 3 $, $ v := s_{1} $, $ v_{r} := s_{1} $};
        \node at (0,-2.2) {$ P := \set{ s_{1} } $};

        \node[anchor=west] at (-1.8,1.4) {\textbf{(4) + (2)}};
        \node at (0,0) {
          \scalebox{0.8} {
            \begin{tikzpicture}[auto,thick]
              \node[state,gray!50,fill=gray!10] (A)                          {$ s_{0} $};
              \node[state,marked]          (B) [below of=A,xshift=-4em] {$ s_{1} $};
              \node[state]          (C) [below of=A,xshift=4em]  {$ s_{2} $};

              \path[->]
              (A) edge[loop above, gray!50] (A)
              (A) edge[out=0,in=75,gray!50] (C)
              (B) edge[gray!50] (A)
              (B) edge[bend left=10] (C)
              (C) edge[bend left=10] (B)   
              (C) edge[gray!50] (A)
              ;
            \end{tikzpicture}
          }
        };
      \end{tikzpicture}
    };

    \node[draw, inner sep=3pt,rounded corners=3] (S8) at (4,-9.2) {
      \begin{tikzpicture}
        \node at (0,-1.8) {$ c := 3 $, $ v := s_{2} $, $ v_{r} := s_{1} $};
        \node at (0,-2.2) {$ P := \set{ s_{1},s_{2} } $};

        \node[anchor=west] at (-1.8,1.4) {\textbf{(3) + (3b)}};
        \node[anchor=west] at (-1.8,0.95) {$ (s_{1}, s_{2}) $};
        \node at (0,0) {
          \scalebox{0.8} {
            \begin{tikzpicture}[auto,thick]
              \node[state,gray!50,fill=gray!10] (A)                          {$ s_{0} $};
              \node[state,marked]          (B) [below of=A,xshift=-4em] {$ s_{1} $};
              \node[state]          (C) [below of=A,xshift=4em]  {$ s_{2} $};

              \path[->]
              (A) edge[loop above, gray!50] (A)
              (A) edge[out=0,in=75,gray!50] (C)
              (B) edge[gray!50] (A)
              (B) edge[bend left=10,ultra thick,red] (C)
              (C) edge[bend left=10] (B)   
              (C) edge[gray!50] (A)
              ;
            \end{tikzpicture}
          }
        };
      \end{tikzpicture}
    };

    \node[draw, inner sep=3pt,rounded corners=3] (S9) at (8,-9.2) {
      \begin{tikzpicture}
        \node at (0,-1.8) {$ c := 4 $, $ v := s_{2} $, $ v_{r} := s_{1} $};
        \node at (0,-2.2) {$ P := \set{ s_{1},s_{2} } $};

        \node[anchor=west] at (-1.8,1.4) {\textbf{(3) + (3a)}};
        \node[anchor=west] at (-1.8,0.95) {$ (s_{2}, s_{1}) $};
        \node at (0,0) {
          \scalebox{0.8} {
            \begin{tikzpicture}[auto,thick]
              \node[state,gray!50,fill=gray!10] (A)                          {$ s_{0} $};
              \node[state,marked]          (B) [below of=A,xshift=-4em] {$ s_{1} $};
              \node[state]          (C) [below of=A,xshift=4em]  {$ s_{2} $};

              \path[->]
              (A) edge[loop above, gray!50] (A)
              (A) edge[out=0,in=75,gray!50] (C)
              (B) edge[gray!50] (A)
              (B) edge[bend left=10,ultra thick,red] (C)
              (C) edge[bend left=10,ultra thick,red] (B)   
              (C) edge[gray!50] (A)
              ;
            \end{tikzpicture}
          }
        };
      \end{tikzpicture}
    };

    \draw ($ (S1.north east) + (-0.7,0) $) edge[out=-90,in=-180] ($ (S1.north east) + (0,-0.7) $);
    \node at ($ (S1.north east) + (-0.3,-0.3) $) {\large 1};

    \draw ($ (S2.north east) + (-0.7,0) $) edge[out=-90,in=-180] ($ (S2.north east) + (0,-0.7) $);
    \node at ($ (S2.north east) + (-0.3,-0.3) $) {\large 2};

    \draw ($ (S3.north east) + (-0.7,0) $) edge[out=-90,in=-180] ($ (S3.north east) + (0,-0.7) $);
    \node at ($ (S3.north east) + (-0.3,-0.3) $) {\large 3};

    \draw ($ (S4.north east) + (-0.7,0) $) edge[out=-90,in=-180] ($ (S4.north east) + (0,-0.7) $);
    \node at ($ (S4.north east) + (-0.3,-0.3) $) {\large 4};

    \draw ($ (S5.north east) + (-0.7,0) $) edge[out=-90,in=-180] ($ (S5.north east) + (0,-0.7) $);
    \node at ($ (S5.north east) + (-0.3,-0.3) $) {\large 5};

    \draw ($ (S6.north east) + (-0.7,0) $) edge[out=-90,in=-180] ($ (S6.north east) + (0,-0.7) $);
    \node at ($ (S6.north east) + (-0.3,-0.3) $) {\large 6};

    \draw ($ (S7.north east) + (-0.7,0) $) edge[out=-90,in=-180] ($ (S7.north east) + (0,-0.7) $);
    \node at ($ (S7.north east) + (-0.3,-0.3) $) {\large 7};

    \draw ($ (S8.north east) + (-0.7,0) $) edge[out=-90,in=-180] ($ (S8.north east) + (0,-0.7) $);
    \node at ($ (S8.north east) + (-0.3,-0.3) $) {\large 8};

    \draw ($ (S9.north east) + (-0.7,0) $) edge[out=-90,in=-180] ($ (S9.north east) + (0,-0.7) $);
    \node at ($ (S9.north east) + (-0.3,-0.3) $) {\large 9};
  \end{tikzpicture}
  \caption{Execution of Tiernan's algorithm for the larger
    Mealy machine on the left in Fig.~\ref{fig:mealymachines}.}
  \label{fig:countexample}
\end{figure}
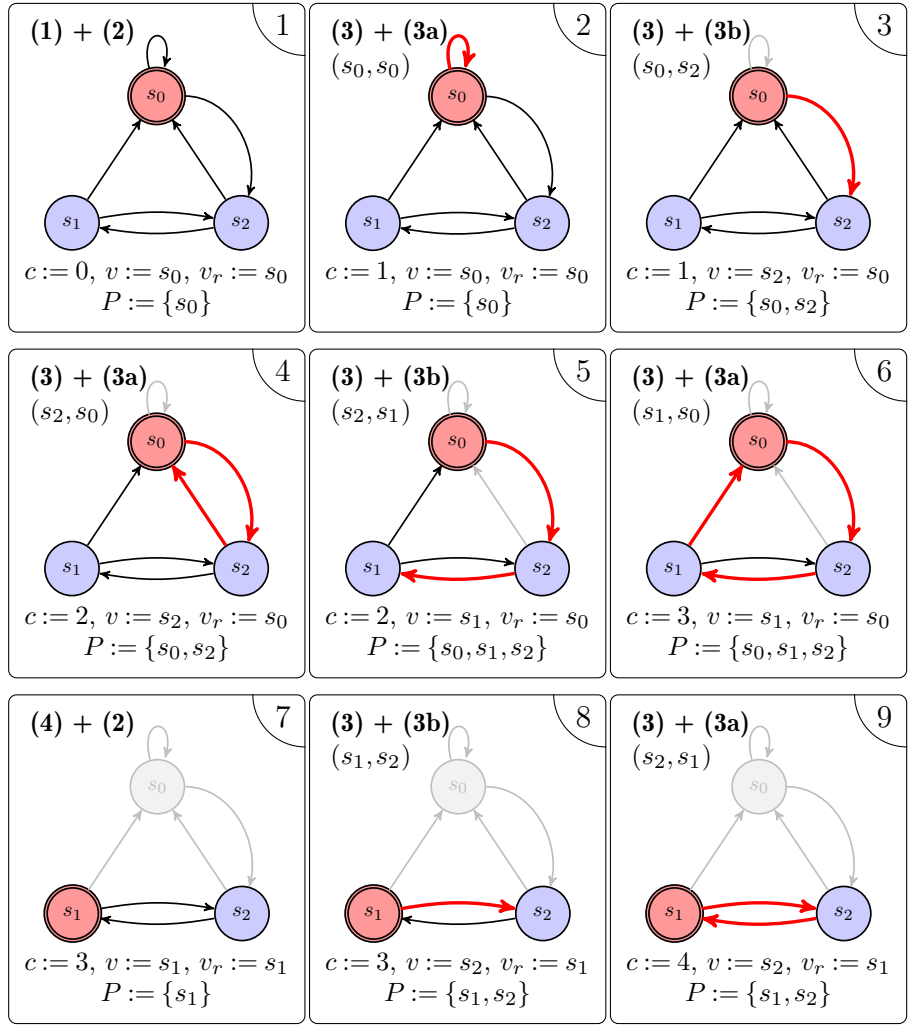

\begin{example}
  To see Tiernan's algorithm in action, we count the number of simple cycles
  of the larger Mealy machine on the left in Fig.~\ref{fig:mealymachines}. The
  execution is shown in Fig.~\ref{fig:countexample}. In this example, we do not need to apply the
  reduction to individual SCCs, because the Mealy machine consists of a single SCC. 
  As result we obtain that the Mealy machine has four
  simple cycles.
\end{example}

\subsection{The Bounded Cycle Synthesis Encoding}

Like in the bounded synthesis approach, we solve the bounded cycle synthesis problem via a reduction to
propositional satisfiability. We extend the constraint system from bounded synthesis with additional
constraints that ensure that the number of cycles, as determined by Tiernan's algorithm, does not exceed the given bound.


We call a tree that witnesses $ m $ cycles in $ G $, all containing
the root~$ r $ of the tree, a witness-tree~$ \tree_{r,m} $ of $ G $.
Formally, a \emph{witness-tree}~$ \tree_{r,m} $ of $ G = (V,E) $ is a labeled
graph $ \tree_{r,m} = ((W,B\cup R),\tau) $, consisting of a graph
$ (W,B \cup R) $ with $ m = \size{R} $ and a labeling function
$ \tau \colon W \rightarrow V $, such that:

\begin{enumerate}

\item The edges are partitioned into blue edges~$ B $ and red
  edges~$ R $.

  \smallskip

  \label{con:witness_tree_first}

\item All red edges lead back to the root:

  \smallskip

  \quad $ R \subseteq W \times \set{ r } $

  \smallskip

  \label{con:witness_tree_red}  
  
\item No blue edges lead back to the root:
  
  \smallskip

  \quad $ B \cap W \times \set{ r } = \emptyset $
  
  \smallskip

  \label{con:witness_noblue}  

\item Each non-root has at least one blue incoming edge:

  \smallskip

  \quad $ \forall w' \in W \setminus \set{ r }.\ \exists w \in W.\ (w,w')
  \in B $ 

  \smallskip

\item Each vertex has at most one blue incoming edge:

  \smallskip

  \quad $ \forall w_{1},w_{2},w \in W.\ (w_{1},w) \in B \wedge (w_{2},w) \in
  B \Rightarrow w_{1} = w_{2} $

  \smallskip

  \label{con:witness_mostblue}  

\item The graph is labeled by an unfolding of $ G $:

  \smallskip
  
  \quad $ \forall w,w' \in B \cup R.\ (\tau(w),\tau(w')) \in E $, 

  \smallskip

\item The unfolding is complete: 

  \smallskip

  \quad
  $ \forall w \in W.\ \forall v' \in V.\ (\tau(w),v') \in E
  \Rightarrow \exists w' \in W.\ (w,w') \in B \cup R \wedge \tau(w') =
  v' $

  \smallskip

  \label{con:witness_tree_completness}

\item Let $ w_{i}, w_{j} \in W $ be two different vertices that appear
  on a path from the root to a leaf in the $ r $-rooted tree
  $ (W,B) $\footnote{Note that the tree property is enforced by
    Conditions~\ref{con:witness_noblue} --
    \ref{con:witness_mostblue}.}. Then the labeling of $ w_{i} $ and
  $ w_{j} $ differs, i.e., $ \tau(v_{i}) \neq \tau(v_{j}) $.

  \smallskip

  \label{con:witness_tree_nodouble}

\item The root of the tree is the same as the corresponding vertex of
  $ G $, i.e., $ \tau(r) = r $.

  \label{con:witness_tree_last}

\end{enumerate}

\begin{lemma}[\cite{FK:2016}]
  Let $ G = (V,E) $ be a graph consisting of a single SCC, $ r \in V $
  be some vertex of $ G $ and $ m $ be the number of cycles of $ G $
  containing $ r $. Then there is a
  witness-tree~$ \tree_{r,m} = ((W,B \cup R),\tau) $ of $ G $ with
  $ \size{W} \leq m \cdot \size{V} $.
  \label{lem:cycles_to_witness}
\end{lemma}

\begin{lemma}[\cite{FK:2016}]
  Let $ G = (V,E) $ be a graph consisting of a single SCC and let
  $ \tree_{r,m} $ be a witness-tree of $ G $. Then there are at most
  $ m $ cycles in $ G $ that contain $ r $.
  \label{lem:witness_to_cycles}
\end{lemma}

\noindent From Lemma~\ref{lem:cycles_to_witness}
and~\ref{lem:witness_to_cycles} we derive that $ \tree_{r,m} $ is a
suitable witness to bound the number of cycles of an
implementation~$ \mealy $. Furthermore, from
Lemma~\ref{lem:cycles_to_witness}, we also obtain an upper bound on the
size of $ \tree_{r,m} $.

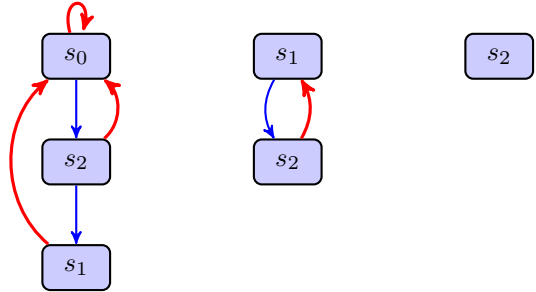
\begin{figure}
  \centering
  \begin{tikzpicture}[thick,node distance=4em]
    \node[state,rectangle,rounded corners=3,minimum height=1.7em] (A) {$ s_{0} $};
    \node[state,rectangle,rounded corners=3,minimum height=1.7em] (B) [below of=A] {$ s_{2} $};
    \node[state,rectangle,rounded corners=3,minimum height=1.7em] (C) [below of=B] {$ s_{1} $};

    \path[->] 
    (A) edge[loop above, very thick, red] (A)
    (A) edge[blue] (B)
    (B) edge[bend right=50, very thick, red] (A)
    (B) edge[blue] (C)
    (C) edge[bend left=50, very thick, red] (A)
    ;

    \node[state,rectangle,rounded corners=3,minimum height=1.7em] (D) [right of=A,xshift=4em] {$ s_{1} $};
    \node[state,rectangle,rounded corners=3,minimum height=1.7em] (E) [below of=D] {$ s_{2} $};

    \path[->] 
    (D) edge[blue, bend right] (E)
    (E) edge[bend right, very thick, red] (D)
    ;

    \node[state,rectangle,rounded corners=3,minimum height=1.7em] (F) [right of=D,xshift=4em] {$ s_{2} $};
  \end{tikzpicture}
  \caption{The forest of witness trees proving prove the overall
    number of four cycles in the larger Mealy machine of
    Fig.~\ref{fig:mealymachines}.}
  \label{fig:witnesstrees}
\end{figure}

\begin{example}
  Figure~\ref{fig:witnesstrees} shows the witness trees for the larger
  Mealy machine on the left of Fig.~\ref{fig:mealymachines}. Each red edge,
  leading back to $ s_{0} $ and $ s_{1} $ on the first tree level,
  captures one cycle of the machine. Thereby, the properties of the
  tree enforce that all cycles are captured by these trees.
\end{example}

\smallskip

We now encode the bound on the number of cycles as a propositional constraint. First,
we construct a simple directed graph~$ G $ out of the
implementation~$ \mealy $.  Then, we guess all the sub-graphs,
obtained from $ G $ via iteratively removing vertices, and split them
into their corresponding SCCs. Finally, we guess the witness-tree for
each such SCC.

In order to keep the encoding compact, we introduce some further
optimizations. First, we do not need to introduce a fresh copy for
each SCC, since the SCC of a vertex is always unique. Thus, it
suffices to guess an annotation for each vertex. Second, we have to guess $ n $ trees~$ \tree_{r_{i},m_i}, i=1 \ldots n$, each
consisting of at most $ m_i \cdot n $ vertices, such that the sum of
all $ m_i $ is equal to the overall number of cycles~$ m $. One possible
solution would be to overestimate each $ m_i $ by $ m $. Another
possibility would be to guess the exact distribution of the cycles
over the different witness-trees~$ \tree_{r_{i},m_i} $. In our encoding, we guess all trees together in a single graph
bounded by $ m \cdot n $. We annotate each vertex with its
corresponding witness-tree~$ \tree_{r_{i},m_i} $. Instead of
bounding the number of red edges separately for each $ \tree_{r_{i},m_i} $ by~$ m_i $,
we just bound the number of all red edges in the whole forest by~$ m
$.
In this way, we not only reduce the size of the encoding, but also avoid
additional constrains that would be needed to sum up the
different witness-tree bounds~$ i $ to $ m $.


\medskip

\noindent Let $ T $ be some ordered set with $ \size{T} = n $ and
$ S = T \times \set{ 1,2,\ldots, m } $. We use $ T $ to denote the
vertices of $ G $ and $ S $ to denote the vertices of the forest of
$ \tree_{r_{i},m_i} $\,s. Further, we use $ M = T \times \set{ 1 } $ to
denote the roots and $ N = S \setminus M $ to denote the non-roots of
the corresponding trees.  We introduce the following Boolean variables:

\begin{itemize}

\item $ \edge{t}{t'} $ for all $ t,t' \in T $, denoting the edges of
  the abstraction of $ \mealy $ to~$ G $.

\item $ \bedge{s}{s'} $ for all $ s \in S $ and
  $ s' \in N $, denoting a blue edge.

\item $ \redge{s}{s'} $ for all $ s \in S $ and $ s' \in M $, denoting a
  red edge.

\item $ \wtreed{s}{i} $ for all $ s \in S $, $ 0 < i \leq \log(n) $,
  denoting the witness-tree for each $ s $. Thereby, each tree is
  referenced by a unique number encoded in binary using a logarithmic
  number of bits. 

\item $ \allowed{s}{t} $ for all $ s \in S $ and $ t \in T $, denoting
  the set of all vertices $ t $, already visited at $ s $, since
  leaving the root of the corresponding witness-tree.

\item $ \rboundd{c}{i} $ for all $ 0 < c \leq m $,
  $ 0 < i \leq \log (n \cdot m) $, denoting an ordered list of all red
  edges, bounding the red edges of the forest.

\item $ \sccd{k}{t}{i} $ for all $ 0 < k \leq n $, $ t \in T, $ and
  $ 0 \leq i < \log n $, denoting the SCC of $ t $ in the $ k $-th
  sub-graph of $ G $. The sub-graphs are obtained 
  by iteratively removing vertices of $ T $, according to the
  pre-defined order. This way, each sub-graph contains exactly all
  vertices that are larger than the root.
 
\end{itemize}
\noindent Note that, by the definition of $ S $, we introduce $ m $
explicit copies for each vertex of~$ G $. This is sufficient, since
each cycle contains each vertex at most once. Thus, the labeling
$ \tau $ of a vertex $ s $ can be directly derived from the first
component of~$ s $.

\medskip

\noindent Given the respective bounded synthesis encoding for the
specification~$ \varphi $ and a bound~$ n $ on the states of the
resulting implementation~$ \mealy $, and a bound $ m $ on the number
of cycles of $ \mealy $, we encode the bounded cycle synthesis problem
as the propositional formula
\begin{equation*}
\curlyF = \curlyF_{BS}(\varphi,n) \ \wedge \ \curlyF_{CS}(n,m) \ \wedge \ 
\curlyF_{\mealy \to G}(\varphi,n) \ \wedge \ \curlyF_{SCC}(n)
\end{equation*}
The constraints of $ \curlyF_{BS}(\varphi,n) $ represent the
bounded synthesis encoding. The constraints of
$ \curlyF_{\mealy \to G}(\varphi,n) $ simplify the representation of
the Mealy machine~$ \mealy $ to $ G $.  The constraints of
$ \curlyF_{CS}(\aut,n,m) $ bound the cycles of the system and are
presented in Table~\ref{tab:constraints}. The constraints of
$ \curlyF_{SCC}(n) $ enforce that each vertex is labeled by a
unique SCC~\cite{FK:2016}.

\begin{table}
  \centering
\caption{Constraints of the SAT formula~$ \curlyF_{CS}(\aut, n, m) $.}
\label{tab:constraints}
\renewcommand{\arraystretch}{1.3}
\scalebox{0.91}{
\begin{tabular}{|>{\centering}m{0.17\textwidth} m{0.43\textwidth} | >{\small}m{0.4\textwidth-12pt} |}
  \hline
  $ \bigwedge\limits_{r \in T} $ & $ \wtree{(r,1)} = r $
  & Roots indicate the witness-tree.
  \\ \hline
  $ \bigwedge\limits_{s \in S,\, (r,1) \in M} $ & $\redge{s}{(r,1)} 
  \rightarrow \wtree{s} = r $
  & Red edges only connect vertices of the current 
    $ \tree_{r_{i},m_i} $. 
  \\ \hline
  $ \bigwedge\limits_{s \in S,\, s' \in N} $ & $ \bedge{s}{s'} \newline
                                               \mbox{\quad} \rightarrow \wtree{s} = \wtree{s'} $ 
  & Blue edges only connect vertices of the
    current $ \tree_{r_{i},m_i} $.
  \\ \hline
  $ \bigwedge\limits_{s' \in N} $ & $ \textit{exactlyOne}(\newline \mbox{\quad} \set{ \bedge{s}{s'} \mid s \in S }\ ) $
  & Every non-root has exactly one blue incoming edge.
  \\ \hline
  $ \bigwedge\limits_{(t,c) \in S,\, r \in T,} $ & $ \redge{(t,c)}{(r,1)} \rightarrow \edge{t}{r} $
  & Red edges are related to the edges of the graph~$ G $.
  \\ \hline
  $ \bigwedge\limits_{(t,c) \in S,\, (t',c') \in N} $ & $ \bedge{(t,c)}{(t',c')} \rightarrow \edge{t}{t'} $  
  & Blue edges are related to the edges of the graph~$ G $.
  \\ \hline
  $ \bigwedge\limits_{\substack{(t,c) \in S,\, r \in T,\\ t \geq r}}  $ &
  $ \edge{t}{r} \wedge \scc{r}{t} = \scc{r}{r} \wedge \newline \wtree{(t,c)} = r \newline \mbox{\quad} \rightarrow \redge{(t,c)}{(r,1)} $ & Every possible red edge must be taken.
  \\ \hline
  $ \bigwedge\limits_{\substack{(t,c) \in S,\, r,t' \in T,\\ t \geq t'}}  $ &
  $ \edge{t}{t'} \wedge \scc{r}{t} = \scc{r}{t'} \wedge \newline \wtree{(t,c)} = r \wedge \allowed{(t,c)}{t'} \newline \mbox{\quad} \rightarrow \bigvee\limits_{0 < c' \leq m} \bedge{(t,c)}{(t',c')} $ & Every possible blue edge must be taken.
  \\ \hline
  $ \bigwedge\limits_{r \in T} $ & $ \bigwedge\limits_{t \leq r} \neg \allowed{(r,1)}{t} \wedge \newline \bigwedge\limits_{t > r} \allowed{(r,1)}{t} $ & 
  Only non-roots of the corresponding sub-graph can be successors of a root.
  \\ \hline
  $ \bigwedge\limits_{(t,c) \in S,\, s \in N} $ & $ \bedge{(t,c)}{s} \newline \mbox{\ \ } \rightarrow \neg \allowed{s}{t} \wedge \newline \mbox{\quad \quad\,} (\allowed{s}{t'} \newline \mbox{\qquad\quad} \leftrightarrow \allowed{(t,c)}{t'}) $ & Every vertex appears at most once on a path from the root to a leaf. 
  \\ \hline

  $ \bigwedge\limits_{s \in S,\, s' \in M} $ & $ \redge{s}{s'} \newline \mbox{\ \ } \rightarrow \bigvee\limits_{0 < c \leq m} \rbound{c} = f(s) $ & The list of red edges is complete. ($ f(s) $ maps each state of $ S $ to a unique number in $ \set{ 1,\ldots,n\cdot m } $)
  \\ \hline
  $ \bigwedge\limits_{0 < c \leq m} $ & $ \rbound{c} < \rbound{c + 1} $ & Red edges are strictly ordered.
  \\ \hline
\end{tabular}
}

\renewcommand{\arraystretch}{1}
\end{table}

\smallskip

\begin{figure}[t]
\centering
\begin{tikzpicture}[auto,semithick]
  \node[state, initial] (A)                         {$ t_{0} $};
  \node[state]          (B) [right of=A,xshift=8em] {$ t_{1} $};

  \path[->]
  (A) edge[bend left=10] node[above] {
        $ * \to \set{ \grant_{1} } $
      } (B)
  (B) edge[bend left=10] node[below] {
        $ * \to \set{ \grant_{2} } $
      } (A)
  ;
\end{tikzpicture}
\caption{The implementation of the arbiter specification with the smallest number of states and cycles.
  \label{fig:minimalsol}}
\end{figure}
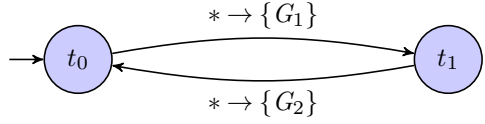

\begin{theorem}[Bounded Cycle Synthesis~\cite{FK:2016}]
  For an LTL formula $\varphi$ and a pair of bounds $ n, m \in \nats $, the propositional 
  formula~$ \curlyF $ is satisfiable if and
  only if there is a Mealy machine~$ \mealy $ with
  $ \size{\mealy} = n $ and \mbox{$ \size{\mathcal{C}(\mealy)} = m $}
  that satisfies $\varphi$.
\end{theorem}

\begin{example}
Using our encoding, we can now 
search for the implementation of the arbiter specification from Example~\ref{example:arbiter} with the smallest number of states and, additionally, smallest number of cycles.
It turns out that neither Mealy machine from
Fig.~\ref{fig:mealymachines} is the minimal solution.
The smallest implementation for the arbiter specification, with respect
to the number of states and cycles is shown in
Fig.~\ref{fig:minimalsol}. The minimal implementation switches the grant at
every time step, completely ignoring the requests. This solution only
requires two states and a single cycle. The solution may not be the
best choice with respect to a possible target application, but it is
definitely the smallest one.
\end{example}

In general, it is not always possible to minimize the two parameters simultaneously.
There are specifications for which the
smallest possible number of states and the smallest possible number of
cycles cannot be realized within a single
solution~\cite{FK:2016}. In such situations it may be helpful to have an explicit
optimization function specified by the user that resolves the trade-off.

\section{Conclusions}

We have studied three different algorithms for the reactive synthesis problem.
The classic game-based synthesis algorithm is input-sensitive in the sense that its performance is asymptotically
optimal in the size of the specification, but it produces implementations that may be larger than necessary.
Bounded synthesis produces implementations with a minimal number of states. Bounded cycle synthesis
additionally minimizes the number of cycles. Bounded synthesis and bounded cycle synthesis belong
to the new class of output-sensitive synthesis algorithms.

%

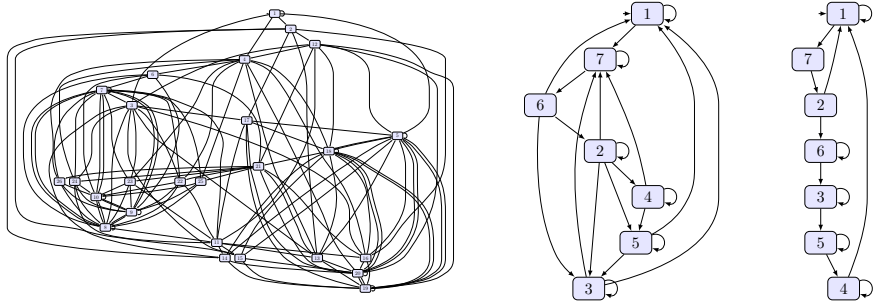
\begin{figure}[t]
\begin{tikzpicture}
  \node at (0,0) {
\scalebox{0.6}{\begin{tikzpicture}[>=latex,line join=bevel,scale=0.4]
\node (1) at (195.0bp,450.0bp) [draw,rounded corners=3,fill=blue!10,inner sep=4pt,minimum width=2em,initial text=,initial left] {1};
  \node (3) at (103.0bp,18.0bp) [draw,rounded corners=3,fill=blue!10,inner sep=4pt,minimum width=2em] {3};
  \node (2) at (121.0bp,234.0bp) [draw,rounded corners=3,fill=blue!10,inner sep=4pt,minimum width=2em] {2};
  \node (5) at (177.0bp,90.0bp) [draw,rounded corners=3,fill=blue!10,inner sep=4pt,minimum width=2em] {5};
  \node (4) at (196.0bp,162.0bp) [draw,rounded corners=3,fill=blue!10,inner sep=4pt,minimum width=2em] {4};
  \node (7) at (121.0bp,378.0bp) [draw,rounded corners=3,fill=blue!10,inner sep=4pt,minimum width=2em] {7};
  \node (6) at (27.0bp,306.0bp) [draw,rounded corners=3,fill=blue!10,inner sep=4pt,minimum width=2em] {6};
  \draw [->] (4) ..controls (228.69bp,177.68bp) and (241.0bp,173.53bp)  .. (241.0bp,162.0bp) .. controls (241.0bp,153.62bp) and (234.5bp,149.14bp)  ..  (4);
  \draw [->] (6) ..controls (58.95bp,281.21bp) and (79.223bp,266.11bp)  .. (2);
  \draw [->] (2) ..controls (116.44bp,178.83bp) and (109.07bp,91.181bp)  .. (3);
  \draw [->] (1) ..controls (227.69bp,465.68bp) and (240.0bp,461.53bp)  .. (240.0bp,450.0bp) .. controls (240.0bp,441.62bp) and (233.5bp,437.14bp)  ..  (1);
  \draw [->] (7) ..controls (89.05bp,353.21bp) and (68.777bp,338.11bp)  ..  (6);
  \draw [->] (2) ..controls (153.69bp,249.68bp) and (166.0bp,245.53bp)  .. (166.0bp,234.0bp) .. controls (166.0bp,225.62bp) and (159.5bp,221.14bp)  ..  (2);
  \draw [->] (7) ..controls (153.69bp,393.68bp) and (166.0bp,389.53bp)  .. (166.0bp,378.0bp) .. controls (166.0bp,369.62bp) and (159.5bp,365.14bp)  .. (7);
  \draw [->] (4) ..controls (189.42bp,198.07bp) and (183.13bp,227.43bp)  .. (175.0bp,252.0bp) .. controls (163.26bp,287.46bp) and (145.44bp,326.77bp)  ..  (7);
  \draw [->] (3) ..controls (183.11bp,36.763bp) and (307.0bp,74.912bp)  .. (307.0bp,161.0bp) .. controls (307.0bp,307.0bp) and (307.0bp,307.0bp)  .. (307.0bp,307.0bp) .. controls (307.0bp,361.47bp) and (255.56bp,407.33bp)  ..  (1);
  \draw [->] (5) ..controls (209.69bp,105.68bp) and (222.0bp,101.53bp)  .. (222.0bp,90.0bp) .. controls (222.0bp,81.622bp) and (215.5bp,77.143bp)  .. (5);
  \draw [->] (4) ..controls (189.25bp,136.14bp) and (186.65bp,126.54bp)  ..  (5);
  \draw [->] (1) ..controls (169.49bp,424.87bp) and (155.22bp,411.37bp)  ..  (7);
  \draw [->] (3) ..controls (91.025bp,74.636bp) and (73.66bp,171.65bp)  .. (85.0bp,252.0bp) .. controls (89.937bp,286.98bp) and (102.28bp,325.89bp)  ..  (7);
  \draw [->] (6) ..controls (43.602bp,342.85bp) and (61.452bp,375.3bp)  .. (85.0bp,396.0bp) .. controls (107.29bp,415.59bp) and (138.1bp,429.61bp)  ..  (1);
  \draw [->] (2) ..controls (121.0bp,276.42bp) and (121.0bp,320.89bp)  ..  (7);
  \draw [->] (3) ..controls (135.69bp,33.675bp) and (148.0bp,29.531bp)  .. (148.0bp,18.0bp) .. controls (148.0bp,9.6218bp) and (141.5bp,5.1433bp)  .. (3);
  \draw [->] (5) ..controls (151.49bp,64.872bp) and (137.22bp,51.369bp)  ..  (3);
  \draw [->] (5) ..controls (215.24bp,110.09bp) and (237.45bp,124.7bp)  .. (250.0bp,144.0bp) .. controls (272.05bp,177.91bp) and (269.0bp,192.55bp)  .. (269.0bp,233.0bp) .. controls (269.0bp,307.0bp) and (269.0bp,307.0bp)  .. (269.0bp,307.0bp) .. controls (269.0bp,353.41bp) and (237.53bp,399.67bp)  ..  (1);
  \draw [->] (2) ..controls (137.13bp,192.09bp) and (155.15bp,146.4bp)  .. (5);
  \draw [->] (6) ..controls (21.294bp,247.93bp) and (16.153bp,146.16bp)  .. (51.0bp,72.0bp) .. controls (57.056bp,59.112bp) and (67.568bp,47.478bp)  .. (3);
  \draw [->] (2) ..controls (146.85bp,208.87bp) and (161.32bp,195.37bp)  .. (4);
\end{tikzpicture}}
};

\node at (2.8,0) {
\scalebox{0.6}{\begin{tikzpicture}[>=latex,line join=bevel,scale=0.4]
\node (1) at (82.0bp,450.0bp) [draw,rounded corners=3,fill=blue!10,inner sep=4pt,minimum width=2em,initial text=,initial left] {1};
  \node (3) at (47.0bp,162.0bp) [draw,rounded corners=3,fill=blue!10,inner sep=4pt,minimum width=2em] {3};
  \node (2) at (47.0bp,306.0bp) [draw,rounded corners=3,fill=blue!10,inner sep=4pt,minimum width=2em] {2};
  \node (5) at (47.0bp,90.0bp) [draw,rounded corners=3,fill=blue!10,inner sep=4pt,minimum width=2em] {5};
  \node (4) at (83.0bp,18.0bp) [draw,rounded corners=3,fill=blue!10,inner sep=4pt,minimum width=2em] {4};
  \node (7) at (27.0bp,378.0bp) [draw,rounded corners=3,fill=blue!10,inner sep=4pt,minimum width=2em] {7};
  \node (6) at (47.0bp,234.0bp) [draw,rounded corners=3,fill=blue!10,inner sep=4pt,minimum width=2em] {6};
  \draw [->] (1) ..controls (114.69bp,465.68bp) and (127.0bp,461.53bp)  .. (127.0bp,450.0bp) .. controls (127.0bp,441.62bp) and (120.5bp,437.14bp)  ..  (1);
  \draw [->] (4) ..controls (115.69bp,33.675bp) and (128.0bp,29.531bp)  .. (128.0bp,18.0bp) .. controls (128.0bp,9.6218bp) and (121.5bp,5.1433bp)  ..  (4);
  \draw [->] (2) ..controls (55.81bp,334.42bp) and (59.887bp,347.9bp)  .. (63.0bp,360.0bp) .. controls (68.28bp,380.53bp) and (73.244bp,403.98bp)  ..  (1);
  \draw [->] (5) ..controls (79.688bp,105.68bp) and (92.0bp,101.53bp)  .. (92.0bp,90.0bp) .. controls (92.0bp,81.622bp) and (85.501bp,77.143bp)  ..  (5);
  \draw [->] (6) ..controls (79.688bp,249.68bp) and (92.0bp,245.53bp)  .. (92.0bp,234.0bp) .. controls (92.0bp,225.62bp) and (85.501bp,221.14bp)  ..  (6);
  \draw [->] (7) ..controls (34.102bp,352.14bp) and (36.846bp,342.54bp)  ..  (2);
  \draw [->] (1) ..controls (62.786bp,424.55bp) and (53.337bp,412.52bp)  ..  (7);
  \draw [->] (4) ..controls (101.09bp,61.849bp) and (120.0bp,114.38bp)  .. (120.0bp,161.0bp) .. controls (120.0bp,307.0bp) and (120.0bp,307.0bp)  .. (120.0bp,307.0bp) .. controls (120.0bp,348.59bp) and (104.6bp,394.78bp)  ..  (1);
  \draw [->] (3) ..controls (79.688bp,177.68bp) and (92.0bp,173.53bp)  .. (92.0bp,162.0bp) .. controls (92.0bp,153.62bp) and (85.501bp,149.14bp)  ..  (3);
  \draw [->] (2) ..controls (47.0bp,279.98bp) and (47.0bp,270.71bp)  ..  (6);
  \draw [->] (5) ..controls (59.712bp,64.283bp) and (65.147bp,53.714bp)  .. (4);
  \draw [->] (3) ..controls (47.0bp,135.98bp) and (47.0bp,126.71bp)  ..  (5);
  \draw [->] (6) ..controls (47.0bp,207.98bp) and (47.0bp,198.71bp)  ..  (3);
\end{tikzpicture}}
};

\node at (-5.2,0) {
\scalebox{0.2}{\begin{tikzpicture}[>=latex,line join=bevel,scale=0.4]
\node (24) at (318.0bp,522.0bp) [draw,rounded corners=3,fill=blue!10,inner sep=4pt,minimum width=2em] {24};
  \node (25) at (911.0bp,522.0bp) [draw,rounded corners=3,fill=blue!10,inner sep=4pt,minimum width=2em] {25};
  \node (26) at (246.0bp,522.0bp) [draw,rounded corners=3,fill=blue!10,inner sep=4pt,minimum width=2em] {26};
  \node (20) at (1651.0bp,90.0bp) [draw,rounded corners=3,fill=blue!10,inner sep=4pt,minimum width=2em] {20};
  \node (21) at (1183.0bp,594.0bp) [draw,rounded corners=3,fill=blue!10,inner sep=4pt,minimum width=2em] {21};
  \node (22) at (814.0bp,522.0bp) [draw,rounded corners=3,fill=blue!10,inner sep=4pt,minimum width=2em] {22};
  \node (23) at (578.0bp,522.0bp) [draw,rounded corners=3,fill=blue!10,inner sep=4pt,minimum width=2em] {23};
  \node (1) at (1260.0bp,1314.0bp) [draw,rounded corners=3,fill=blue!10,inner sep=4pt,minimum width=2em,initial text=,initial left] {1};
  \node (3) at (587.0bp,882.0bp) [draw,rounded corners=3,fill=blue!10,inner sep=4pt,minimum width=2em] {3};
  \node (2) at (1335.0bp,1242.0bp) [draw,rounded corners=3,fill=blue!10,inner sep=4pt,minimum width=2em] {2};
  \node (5) at (1838.0bp,738.0bp) [draw,rounded corners=3,fill=blue!10,inner sep=4pt,minimum width=2em] {5};
  \node (4) at (1118.0bp,1098.0bp) [draw,rounded corners=3,fill=blue!10,inner sep=4pt,minimum width=2em] {4};
  \node (7) at (445.0bp,954.0bp) [draw,rounded corners=3,fill=blue!10,inner sep=4pt,minimum width=2em] {7};
  \node (6) at (685.0bp,1026.0bp) [draw,rounded corners=3,fill=blue!10,inner sep=4pt,minimum width=2em] {6};
  \node (9) at (585.0bp,378.0bp) [draw,rounded corners=3,fill=blue!10,inner sep=4pt,minimum width=2em] {9};
  \node (8) at (462.0bp,306.0bp) [draw,rounded corners=3,fill=blue!10,inner sep=4pt,minimum width=2em] {8};
  \node (11) at (987.0bp,234.0bp) [draw,rounded corners=3,fill=blue!10,inner sep=4pt,minimum width=2em] {11};
  \node (10) at (419.0bp,450.0bp) [draw,rounded corners=3,fill=blue!10,inner sep=4pt,minimum width=2em] {10};
  \node (13) at (1460.0bp,162.0bp) [draw,rounded corners=3,fill=blue!10,inner sep=4pt,minimum width=2em] {13};
  \node (12) at (1449.0bp,1170.0bp) [draw,rounded corners=3,fill=blue!10,inner sep=4pt,minimum width=2em] {12};
  \node (15) at (1097.0bp,162.0bp) [draw,rounded corners=3,fill=blue!10,inner sep=4pt,minimum width=2em] {15};
  \node (14) at (1025.0bp,162.0bp) [draw,rounded corners=3,fill=blue!10,inner sep=4pt,minimum width=2em] {14};
  \node (17) at (1128.0bp,810.0bp) [draw,rounded corners=3,fill=blue!10,inner sep=4pt,minimum width=2em] {17};
  \node (16) at (1688.0bp,162.0bp) [draw,rounded corners=3,fill=blue!10,inner sep=4pt,minimum width=2em] {16};
  \node (19) at (1687.0bp,18.0bp) [draw,rounded corners=3,fill=blue!10,inner sep=4pt,minimum width=2em] {19};
  \node (18) at (1515.0bp,666.0bp) [draw,rounded corners=3,fill=blue!10,inner sep=4pt,minimum width=2em] {18};
  \draw [->] (25) ..controls (862.56bp,490.26bp) and (804.09bp,454.97bp)  .. (751.0bp,432.0bp) .. controls (706.97bp,412.95bp) and (653.78bp,397.16bp)  .. (9);
  \draw [->] (21) ..controls (1060.1bp,578.78bp) and (726.17bp,540.14bp)  .. (23);
  \draw [->] (2) ..controls (1295.7bp,1159.3bp) and (1178.8bp,916.58bp)  .. (17);
  \draw [->] (20) ..controls (1657.4bp,63.224bp) and (1662.9bp,52.281bp)  .. (19);
  \draw [->] (24) ..controls (324.32bp,485.38bp) and (331.68bp,455.07bp)  .. (345.0bp,432.0bp) .. controls (369.51bp,389.54bp) and (411.0bp,349.9bp)  ..  (8);
  \draw [->] (4) ..controls (971.27bp,1096.6bp) and (491.37bp,1092.2bp)  .. (350.0bp,1044.0bp) .. controls (295.4bp,1025.4bp) and (271.88bp,1021.3bp)  .. (242.0bp,972.0bp) .. controls (175.43bp,862.06bp) and (218.68bp,812.49bp)  .. (216.0bp,684.0bp) .. controls (215.0bp,635.94bp) and (209.91bp,622.53bp)  .. (222.0bp,576.0bp) .. controls (224.44bp,566.62bp) and (228.41bp,556.83bp)  .. (26);
  \draw [->] (3) ..controls (577.71bp,837.52bp) and (568.0bp,784.52bp)  .. (568.0bp,739.0bp) .. controls (568.0bp,739.0bp) and (568.0bp,739.0bp)  .. (568.0bp,665.0bp) .. controls (568.0bp,533.85bp) and (835.13bp,277.44bp)  .. (951.0bp,216.0bp) .. controls (996.45bp,191.9bp) and (1013.2bp,199.02bp)  .. (1061.0bp,180.0bp) .. controls (1062.8bp,179.27bp) and (1064.7bp,178.49bp)  .. (15);
  \draw [->] (6) ..controls (596.92bp,1017.1bp) and (448.76bp,1000.9bp)  .. (409.0bp,972.0bp) .. controls (372.48bp,945.45bp) and (360.0bp,928.15bp)  .. (360.0bp,883.0bp) .. controls (360.0bp,883.0bp) and (360.0bp,883.0bp)  .. (360.0bp,665.0bp) .. controls (360.0bp,594.83bp) and (389.84bp,516.23bp)  ..  (10);
  \draw [->] (26) ..controls (275.6bp,485.69bp) and (308.82bp,450.6bp)  .. (345.0bp,432.0bp) .. controls (411.29bp,397.92bp) and (498.86bp,385.67bp)  .. (9);
  \draw [->] (13) ..controls (1432.7bp,186.62bp) and (1418.1bp,201.08bp)  .. (1408.0bp,216.0bp) .. controls (1378.5bp,259.44bp) and (1382.9bp,276.75bp)  .. (1360.0bp,324.0bp) .. controls (1312.6bp,421.65bp) and (1314.1bp,455.49bp)  .. (1246.0bp,540.0bp) .. controls (1235.5bp,553.02bp) and (1221.4bp,565.21bp)  .. (21);
  \draw [->] (20) ..controls (1725.6bp,108.1bp) and (1831.9bp,143.32bp)  .. (1881.0bp,216.0bp) .. controls (1958.0bp,329.87bp) and (1932.0bp,383.56bp)  .. (1932.0bp,521.0bp) .. controls (1932.0bp,595.0bp) and (1932.0bp,595.0bp)  .. (1932.0bp,595.0bp) .. controls (1932.0bp,644.95bp) and (1892.8bp,692.54bp)  .. (5);
  \draw [->] (16) ..controls (1696.8bp,190.26bp) and (1700.4bp,203.73bp)  .. (1702.0bp,216.0bp) .. controls (1712.3bp,295.33bp) and (1712.3bp,316.66bp)  .. (1702.0bp,396.0bp) .. controls (1693.5bp,461.64bp) and (1699.5bp,484.11bp)  .. (1664.0bp,540.0bp) .. controls (1633.3bp,588.34bp) and (1577.9bp,627.48bp)  ..  (18);
  \draw [->] (2) ..controls (1139.9bp,1239.1bp) and (237.38bp,1228.4bp)  .. (122.0bp,1188.0bp) .. controls (58.657bp,1165.8bp) and (0.0000bp,1166.1bp)  .. (0.0bp,1099.0bp) .. controls (0.0bp,1099.0bp) and (0.0bp,1099.0bp)  .. (0.0bp,305.0bp) .. controls (0.0bp,203.17bp) and (804.03bp,170.33bp)  ..  (14);
  \draw [->] (12) ..controls (1535.7bp,1162.8bp) and (1684.8bp,1149.0bp)  .. (1805.0bp,1116.0bp) .. controls (1886.5bp,1093.6bp) and (1907.3bp,1085.4bp)  .. (1981.0bp,1044.0bp) .. controls (2039.2bp,1011.3bp) and (2102.0bp,1021.8bp)  .. (2102.0bp,955.0bp) .. controls (2102.0bp,955.0bp) and (2102.0bp,955.0bp)  .. (2102.0bp,161.0bp) .. controls (2102.0bp,111.22bp) and (2076.8bp,97.378bp)  .. (2034.0bp,72.0bp) .. controls (1982.1bp,41.214bp) and (1802.6bp,26.331bp)  .. (19);
  \draw [->] (26) ..controls (273.2bp,473.35bp) and (316.16bp,404.38bp)  .. (367.0bp,360.0bp) .. controls (385.95bp,343.46bp) and (410.9bp,329.78bp)  ..  (8);
  \draw [->] (7) ..controls (436.0bp,909.87bp) and (436.0bp,856.36bp)  .. (436.0bp,811.0bp) .. controls (436.0bp,811.0bp) and (436.0bp,811.0bp)  .. (436.0bp,593.0bp) .. controls (436.0bp,552.55bp) and (425.67bp,506.67bp)  .. (10);
  \draw [->] (5) ..controls (1871.0bp,697.0bp) and (1914.0bp,647.29bp)  .. (1914.0bp,595.0bp) .. controls (1914.0bp,595.0bp) and (1914.0bp,595.0bp)  .. (1914.0bp,521.0bp) .. controls (1914.0bp,383.56bp) and (1940.0bp,329.87bp)  .. (1863.0bp,216.0bp) .. controls (1819.9bp,152.24bp) and (1732.8bp,117.31bp)  ..  (20);
  \draw [->] (21) ..controls (1109.1bp,573.99bp) and (1001.4bp,546.27bp)  .. (25);
  \draw [->] (19) ..controls (1680.5bp,44.873bp) and (1675.1bp,55.8bp)  .. (20);
  \draw [->] (10) ..controls (467.77bp,425.37bp) and (518.42bp,403.99bp)  .. (9);
  \draw [->] (21) ..controls (1103.5bp,581.9bp) and (978.49bp,563.26bp)  .. (875.0bp,540.0bp) .. controls (866.21bp,538.02bp) and (856.8bp,535.57bp)  ..  (22);
  \draw [->] (24) ..controls (355.77bp,561.26bp) and (398.0bp,612.37bp)  .. (398.0bp,665.0bp) .. controls (398.0bp,739.0bp) and (398.0bp,739.0bp)  .. (398.0bp,739.0bp) .. controls (398.0bp,815.31bp) and (497.69bp,855.68bp)  ..  (3);
  \draw [->] (26) ..controls (229.94bp,579.04bp) and (207.12bp,678.44bp)  .. (232.0bp,756.0bp) .. controls (256.04bp,830.93bp) and (274.79bp,849.34bp)  .. (335.0bp,900.0bp) .. controls (357.71bp,919.11bp) and (388.48bp,933.17bp)  .. (7);
  \draw [->] (10) ..controls (440.42bp,493.12bp) and (454.0bp,546.63bp)  .. (454.0bp,593.0bp) .. controls (454.0bp,811.0bp) and (454.0bp,811.0bp)  .. (454.0bp,811.0bp) .. controls (454.0bp,851.22bp) and (454.0bp,897.85bp)  .. (7);
  \draw [->] (14) ..controls (1059.6bp,125.55bp) and (1100.2bp,88.681bp)  .. (1143.0bp,72.0bp) .. controls (1235.8bp,35.809bp) and (1542.5bp,23.327bp)  .. (19);
  \draw [->] (10) ..controls (424.47bp,407.19bp) and (438.07bp,361.45bp)  .. (8);
  \draw [->] (7) ..controls (477.69bp,969.68bp) and (490.0bp,965.53bp)  .. (490.0bp,954.0bp) .. controls (490.0bp,945.62bp) and (483.5bp,941.14bp)  .. (7);
  \draw [->] (21) ..controls (1024.4bp,591.57bp) and (453.56bp,583.8bp)  .. (282.0bp,540.0bp) .. controls (279.82bp,539.44bp) and (277.62bp,538.76bp)  ..  (26);
  \draw [->] (20) ..controls (1683.7bp,105.68bp) and (1696.0bp,101.53bp)  .. (1696.0bp,90.0bp) .. controls (1696.0bp,81.622bp) and (1689.5bp,77.143bp)  .. (20);
  \draw [->] (11) ..controls (1092.7bp,217.36bp) and (1332.2bp,181.92bp)  .. (13);
  \draw [->] (10) ..controls (451.69bp,465.68bp) and (464.0bp,461.53bp)  .. (464.0bp,450.0bp) .. controls (464.0bp,441.62bp) and (457.5bp,437.14bp)  .. (10);
  \draw [->] (24) ..controls (319.96bp,566.7bp) and (322.0bp,619.87bp)  .. (322.0bp,665.0bp) .. controls (322.0bp,811.0bp) and (322.0bp,811.0bp)  .. (322.0bp,811.0bp) .. controls (322.0bp,868.13bp) and (379.58bp,913.56bp)  ..  (7);
  \draw [->] (16) ..controls (1677.3bp,227.58bp) and (1650.8bp,365.32bp)  .. (1599.0bp,468.0bp) .. controls (1555.7bp,553.84bp) and (1530.9bp,567.06bp)  .. (1479.0bp,648.0bp) .. controls (1423.2bp,735.02bp) and (1312.6bp,975.69bp)  .. (1235.0bp,1044.0bp) .. controls (1211.2bp,1065.0bp) and (1177.7bp,1079.0bp)  .. (4);
  \draw [->] (8) ..controls (476.67bp,350.14bp) and (492.0bp,402.89bp)  .. (492.0bp,449.0bp) .. controls (492.0bp,523.0bp) and (492.0bp,523.0bp)  .. (492.0bp,523.0bp) .. controls (492.0bp,661.33bp) and (502.86bp,699.47bp)  .. (554.0bp,828.0bp) .. controls (557.99bp,838.02bp) and (563.86bp,848.25bp)  .. (3);
  \draw [->] (6) ..controls (617.72bp,1005.4bp) and (528.39bp,979.32bp)  .. (7);
  \draw [->] (11) ..controls (1000.4bp,208.28bp) and (1006.2bp,197.71bp)  .. (14);
  \draw [->] (1) ..controls (1285.9bp,1288.9bp) and (1300.3bp,1275.4bp)  .. (2);
  \draw [->] (16) ..controls (1751.3bp,213.31bp) and (1876.0bp,325.11bp)  .. (1876.0bp,449.0bp) .. controls (1876.0bp,595.0bp) and (1876.0bp,595.0bp)  .. (1876.0bp,595.0bp) .. controls (1876.0bp,636.59bp) and (1860.6bp,682.78bp)  .. (5);
  \draw [->] (3) ..controls (615.18bp,834.17bp) and (658.11bp,767.21bp)  .. (705.0bp,720.0bp) .. controls (706.03bp,718.96bp) and (1381.8bp,216.85bp)  .. (1383.0bp,216.0bp) .. controls (1399.4bp,204.27bp) and (1418.0bp,191.44bp)  ..  (13);
  \draw [->] (20) ..controls (1695.3bp,106.18bp) and (1721.5bp,121.4bp)  .. (1733.0bp,144.0bp) .. controls (1798.5bp,272.43bp) and (1748.1bp,327.03bp)  .. (1718.0bp,468.0bp) .. controls (1702.7bp,539.6bp) and (1695.7bp,566.88bp)  .. (1638.0bp,612.0bp) .. controls (1611.6bp,632.63bp) and (1575.9bp,647.57bp)  .. (18);
  \draw [->] (11) ..controls (1023.7bp,209.67bp) and (1049.8bp,193.06bp)  ..  (15);
  \draw [->] (4) ..controls (1083.0bp,999.84bp) and (956.24bp,648.44bp)  .. (25);
  \draw [->] (3) ..controls (524.29bp,873.33bp) and (458.55bp,860.4bp)  .. (413.0bp,828.0bp) .. controls (274.52bp,729.51bp) and (156.5bp,665.29bp)  .. (210.0bp,504.0bp) .. controls (247.69bp,390.35bp) and (269.53bp,351.56bp)  .. (371.0bp,288.0bp) .. controls (476.32bp,222.03bp) and (866.06bp,178.7bp)  .. (14);
  \draw [->] (18) ..controls (1430.8bp,647.26bp) and (1286.2bp,616.76bp)  .. (21);
  \draw [->] (16) ..controls (1698.7bp,133.88bp) and (1703.1bp,120.41bp)  .. (1705.0bp,108.0bp) .. controls (1707.5bp,92.192bp) and (1707.6bp,87.786bp)  .. (1705.0bp,72.0bp) .. controls (1703.5bp,63.139bp) and (1700.8bp,53.768bp)  .. (19);
  \draw [->] (25) ..controls (875.97bp,510.16bp) and (862.4bp,506.48bp)  .. (850.0bp,504.0bp) .. controls (705.9bp,475.23bp) and (531.38bp,459.54bp)  .. (10);
  \draw [->] (6) ..controls (539.15bp,1023.2bp) and (76.0bp,999.77bp)  .. (76.0bp,739.0bp) .. controls (76.0bp,739.0bp) and (76.0bp,739.0bp)  .. (76.0bp,665.0bp) .. controls (76.0bp,502.66bp) and (133.39bp,435.71bp)  .. (277.0bp,360.0bp) .. controls (325.22bp,334.58bp) and (387.12bp,320.03bp)  ..  (8);
  \draw [->] (17) ..controls (1133.3bp,739.04bp) and (1145.5bp,581.19bp)  .. (1147.0bp,576.0bp) .. controls (1208.7bp,365.19bp) and (1204.0bp,265.41bp)  .. (1387.0bp,144.0bp) .. controls (1423.9bp,119.52bp) and (1550.8bp,102.25bp)  .. (20);
  \draw [->] (14) ..controls (1050.4bp,147.54bp) and (1055.8bp,145.41bp)  .. (1061.0bp,144.0bp) .. controls (1166.2bp,115.42bp) and (1501.5bp,97.839bp)  .. (20);
  \draw [->] (24) ..controls (317.26bp,485.61bp) and (318.44bp,456.13bp)  .. (326.0bp,432.0bp) .. controls (349.3bp,357.65bp) and (357.22bp,324.62bp)  .. (426.0bp,288.0bp) .. controls (517.09bp,239.5bp) and (839.06bp,234.94bp)  ..  (11);
  \draw [->] (15) ..controls (1126.7bp,124.85bp) and (1160.9bp,88.689bp)  .. (1199.0bp,72.0bp) .. controls (1280.0bp,36.539bp) and (1550.1bp,23.747bp)  ..  (19);
  \draw [->] (13) ..controls (1519.0bp,124.08bp) and (1612.4bp,65.658bp)  .. (19);
  \draw [->] (2) ..controls (1121.7bp,1240.4bp) and (38.0bp,1231.5bp)  .. (38.0bp,1099.0bp) .. controls (38.0bp,1099.0bp) and (38.0bp,1099.0bp)  .. (38.0bp,449.0bp) .. controls (38.0bp,284.61bp) and (214.09bp,337.02bp)  .. (371.0bp,288.0bp) .. controls (519.14bp,241.72bp) and (912.65bp,225.6bp)  .. (1061.0bp,180.0bp) .. controls (1062.9bp,179.41bp) and (1064.9bp,178.73bp)  ..  (15);
  \draw [->] (23) ..controls (561.51bp,474.15bp) and (536.74bp,409.39bp)  .. (506.0bp,360.0bp) .. controls (499.26bp,349.16bp) and (490.49bp,338.15bp)  ..  (8);
  \draw [->] (3) ..controls (702.18bp,866.1bp) and (989.25bp,828.95bp)  ..  (17);
  \draw [->] (1) ..controls (1225.0bp,1260.2bp) and (1163.4bp,1167.4bp)  ..  (4);
  \draw [->] (22) ..controls (782.17bp,486.85bp) and (748.29bp,453.43bp)  .. (714.0bp,432.0bp) .. controls (683.96bp,413.23bp) and (646.12bp,398.6bp)  ..  (9);
  \draw [->] (9) ..controls (637.94bp,451.97bp) and (753.18bp,663.87bp)  .. (693.0bp,828.0bp) .. controls (678.56bp,867.38bp) and (667.11bp,877.06bp)  .. (632.0bp,900.0bp) .. controls (585.79bp,930.19bp) and (521.78bp,943.7bp)  .. (7);
  \draw [->] (21) ..controls (1193.3bp,652.62bp) and (1205.8bp,757.21bp)  .. (1164.0bp,828.0bp) .. controls (1070.0bp,987.37bp) and (817.6bp,1017.8bp)  ..  (6);
  \draw [->] (12) ..controls (1403.9bp,1071.7bp) and (1240.4bp,718.96bp)  .. (21);
  \draw [->] (18) ..controls (1487.6bp,743.11bp) and (1407.5bp,942.7bp)  .. (1273.0bp,1044.0bp) .. controls (1237.5bp,1070.7bp) and (1187.5bp,1084.6bp)  .. (4);
  \draw [->] (24) ..controls (335.81bp,484.34bp) and (355.81bp,450.54bp)  .. (383.0bp,432.0bp) .. controls (433.51bp,397.57bp) and (505.1bp,385.5bp)  .. (9);
  \draw [->] (18) ..controls (1547.7bp,681.68bp) and (1560.0bp,677.53bp)  .. (1560.0bp,666.0bp) .. controls (1560.0bp,657.62bp) and (1553.5bp,653.14bp)  ..  (18);
  \draw [->] (13) ..controls (1516.7bp,140.21bp) and (1579.6bp,117.15bp)  ..  (20);
  \draw [->] (16) ..controls (1674.9bp,136.28bp) and (1669.3bp,125.71bp)  ..  (20);
  \draw [->] (22) ..controls (794.17bp,485.82bp) and (774.12bp,454.11bp)  .. (751.0bp,432.0bp) .. controls (708.23bp,391.1bp) and (692.76bp,384.73bp)  .. (639.0bp,360.0bp) .. controls (592.05bp,338.4bp) and (534.1bp,323.0bp)  ..  (8);
  \draw [->] (19) ..controls (1719.7bp,33.675bp) and (1732.0bp,29.531bp)  .. (1732.0bp,18.0bp) .. controls (1732.0bp,9.6218bp) and (1725.5bp,5.1433bp)  .. (19);
  \draw [->] (2) ..controls (1372.6bp,1217.9bp) and (1400.2bp,1201.0bp)  .. (12);
  \draw [->] (4) ..controls (1062.7bp,1085.7bp) and (1013.4bp,1071.4bp)  .. (980.0bp,1044.0bp) .. controls (906.75bp,983.98bp) and (922.01bp,940.35bp)  .. (866.0bp,864.0bp) .. controls (773.35bp,737.71bp) and (647.47bp,598.27bp)  .. (23);
  \draw [->] (8) ..controls (411.6bp,322.57bp) and (369.24bp,337.24bp)  .. (338.0bp,360.0bp) .. controls (205.79bp,456.33bp) and (132.0bp,501.42bp)  .. (132.0bp,665.0bp) .. controls (132.0bp,739.0bp) and (132.0bp,739.0bp)  .. (132.0bp,739.0bp) .. controls (132.0bp,872.89bp) and (327.49bp,928.05bp)  .. (7);
  \draw [->] (18) ..controls (1553.4bp,676.05bp) and (1569.0bp,680.59bp)  .. (1583.0bp,684.0bp) .. controls (1661.0bp,703.04bp) and (1753.2bp,720.39bp)  ..  (5);
  \draw [->] (20) ..controls (1578.3bp,91.277bp) and (1484.8bp,98.608bp)  .. (1424.0bp,144.0bp) .. controls (1347.6bp,201.02bp) and (1229.1bp,480.13bp)  .. (21);
  \draw [->] (23) ..controls (591.69bp,566.21bp) and (606.0bp,619.02bp)  .. (606.0bp,665.0bp) .. controls (606.0bp,739.0bp) and (606.0bp,739.0bp)  .. (606.0bp,739.0bp) .. controls (606.0bp,779.36bp) and (598.37bp,825.61bp)  ..  (3);
  \draw [->] (12) ..controls (1365.1bp,1151.3bp) and (1220.9bp,1120.8bp)  ..  (4);
  \draw [->] (9) ..controls (617.69bp,393.68bp) and (630.0bp,389.53bp)  .. (630.0bp,378.0bp) .. controls (630.0bp,369.62bp) and (623.5bp,365.14bp)  .. (9);
  \draw [->] (8) ..controls (456.54bp,348.81bp) and (442.85bp,394.85bp)  .. (10);
  \draw [->] (1) ..controls (1140.9bp,1292.7bp) and (822.6bp,1226.0bp)  .. (649.0bp,1044.0bp) .. controls (613.03bp,1006.3bp) and (597.31bp,945.46bp)  ..  (3);
  \draw [->] (9) ..controls (541.96bp,356.44bp) and (510.4bp,338.7bp)  .. (8);
  \draw [->] (25) ..controls (907.39bp,600.73bp) and (888.56bp,803.97bp)  .. (772.0bp,900.0bp) .. controls (728.28bp,936.02bp) and (558.23bp,947.99bp)  .. (7);
  \draw [->] (17) ..controls (1123.2bp,715.37bp) and (1117.7bp,402.59bp)  .. (1262.0bp,216.0bp) .. controls (1289.5bp,180.45bp) and (1295.1bp,168.1bp)  .. (1333.0bp,144.0bp) .. controls (1438.5bp,76.991bp) and (1584.5bp,40.237bp)  .. (19);
  \draw [->] (12) ..controls (1542.0bp,1159.6bp) and (1706.0bp,1131.2bp)  .. (1706.0bp,1027.0bp) .. controls (1706.0bp,1027.0bp) and (1706.0bp,1027.0bp)  .. (1706.0bp,881.0bp) .. controls (1706.0bp,821.48bp) and (1769.0bp,776.44bp)  .. (5);
  \draw [->] (11) ..controls (999.56bp,317.95bp) and (1039.4bp,563.04bp)  .. (1104.0bp,756.0bp) .. controls (1107.1bp,765.19bp) and (1111.2bp,774.93bp)  .. (17);
  \draw [->] (20) ..controls (1617.0bp,128.42bp) and (1578.6bp,173.17bp)  .. (1553.0bp,216.0bp) .. controls (1508.1bp,290.97bp) and (1511.2bp,316.93bp)  .. (1474.0bp,396.0bp) .. controls (1337.0bp,687.61bp) and (1302.6bp,760.8bp)  .. (1149.0bp,1044.0bp) .. controls (1143.9bp,1053.3bp) and (1138.1bp,1063.4bp)  ..  (4);
  \draw [->] (1) ..controls (1292.7bp,1329.7bp) and (1305.0bp,1325.5bp)  .. (1305.0bp,1314.0bp) .. controls (1305.0bp,1305.6bp) and (1298.5bp,1301.1bp)  ..  (1);
  \draw [->] (7) ..controls (506.16bp,943.19bp) and (568.65bp,929.63bp)  .. (614.0bp,900.0bp) .. controls (649.11bp,877.06bp) and (660.56bp,867.38bp)  .. (675.0bp,828.0bp) .. controls (731.95bp,672.68bp) and (631.8bp,474.56bp)  ..  (9);
  \draw [->] (4) ..controls (943.76bp,1087.1bp) and (246.0bp,1042.8bp)  .. (246.0bp,955.0bp) .. controls (246.0bp,955.0bp) and (246.0bp,955.0bp)  .. (246.0bp,809.0bp) .. controls (246.0bp,710.32bp) and (286.81bp,598.13bp)  ..  (24);
  \draw [->] (23) ..controls (528.48bp,499.2bp) and (481.58bp,478.55bp)  ..  (10);
  \draw [->] (22) ..controls (841.07bp,497.25bp) and (855.65bp,482.78bp)  .. (866.0bp,468.0bp) .. controls (916.85bp,395.37bp) and (908.36bp,365.18bp)  .. (952.0bp,288.0bp) .. controls (957.46bp,278.35bp) and (964.02bp,268.11bp)  ..  (11);
  \draw [->] (9) ..controls (536.19bp,402.65bp) and (485.47bp,424.06bp)  ..  (10);
  \draw [->] (18) ..controls (1560.0bp,646.71bp) and (1594.4bp,632.04bp)  .. (1620.0bp,612.0bp) .. controls (1677.7bp,566.88bp) and (1684.7bp,539.6bp)  .. (1700.0bp,468.0bp) .. controls (1730.1bp,327.03bp) and (1780.5bp,272.43bp)  .. (1715.0bp,144.0bp) .. controls (1707.1bp,128.55bp) and (1692.4bp,116.55bp)  .. (20);
  \draw [->] (5) ..controls (1873.7bp,715.92bp) and (1896.7bp,701.92bp)  .. (1912.0bp,684.0bp) .. controls (2008.3bp,571.55bp) and (2046.0bp,527.06bp)  .. (2046.0bp,379.0bp) .. controls (2046.0bp,379.0bp) and (2046.0bp,379.0bp)  .. (2046.0bp,161.0bp) .. controls (2046.0bp,111.22bp) and (2020.5bp,97.875bp)  .. (1978.0bp,72.0bp) .. controls (1935.3bp,46.035bp) and (1792.3bp,29.531bp)  ..  (19);
  \draw [->] (3) ..controls (669.57bp,935.13bp) and (884.42bp,1064.4bp)  .. (1082.0bp,1116.0bp) .. controls (1200.2bp,1146.9bp) and (1344.9bp,1161.1bp)  ..  (12);
  \draw [->] (11) ..controls (1055.3bp,272.4bp) and (1175.1bp,343.82bp)  .. (1248.0bp,432.0bp) .. controls (1335.2bp,537.53bp) and (1449.0bp,733.18bp)  .. (1449.0bp,881.0bp) .. controls (1449.0bp,1027.0bp) and (1449.0bp,1027.0bp)  .. (1449.0bp,1027.0bp) .. controls (1449.0bp,1067.0bp) and (1449.0bp,1113.3bp)  ..  (12);
  \draw [->] (15) ..controls (1175.1bp,220.35bp) and (1382.4bp,373.18bp)  .. (1550.0bp,504.0bp) .. controls (1649.0bp,581.25bp) and (1764.5bp,676.17bp)  .. (5);
  \draw [->] (24) ..controls (351.76bp,497.6bp) and (374.72bp,481.69bp)  .. (10);
  \draw [->] (7) ..controls (340.06bp,935.92bp) and (114.0bp,882.98bp)  .. (114.0bp,739.0bp) .. controls (114.0bp,739.0bp) and (114.0bp,739.0bp)  .. (114.0bp,665.0bp) .. controls (114.0bp,501.42bp) and (187.79bp,456.33bp)  .. (320.0bp,360.0bp) .. controls (352.04bp,336.66bp) and (395.77bp,321.83bp)  .. (8);
  \draw [->] (15) ..controls (1116.2bp,210.35bp) and (1145.0bp,275.59bp)  .. (1180.0bp,324.0bp) .. controls (1278.8bp,460.51bp) and (1431.6bp,594.96bp)  .. (18);
  \draw [->] (13) ..controls (1494.4bp,223.76bp) and (1560.1bp,353.61bp)  .. (1550.0bp,468.0bp) .. controls (1544.6bp,529.26bp) and (1529.9bp,600.0bp)  .. (18);
  \draw [->] (2) ..controls (1493.3bp,1227.0bp) and (2064.0bp,1172.0bp)  .. (2064.0bp,1099.0bp) .. controls (2064.0bp,1099.0bp) and (2064.0bp,1099.0bp)  .. (2064.0bp,665.0bp) .. controls (2064.0bp,469.87bp) and (1951.1bp,447.68bp)  .. (1839.0bp,288.0bp) .. controls (1815.2bp,254.14bp) and (1811.5bp,242.76bp)  .. (1780.0bp,216.0bp) .. controls (1761.5bp,200.29bp) and (1737.7bp,186.71bp)  .. (16);
  \draw [->] (3) ..controls (768.36bp,834.87bp) and (1612.0bp,609.1bp)  .. (1612.0bp,379.0bp) .. controls (1612.0bp,379.0bp) and (1612.0bp,379.0bp)  .. (1612.0bp,305.0bp) .. controls (1612.0bp,258.12bp) and (1644.6bp,211.74bp)  ..  (16);
  \draw [->] (17) ..controls (1263.3bp,795.66bp) and (1674.9bp,755.08bp)  .. (5);
  \draw [->] (14) ..controls (1115.1bp,240.21bp) and (1434.1bp,508.81bp)  .. (1726.0bp,684.0bp) .. controls (1752.0bp,699.6bp) and (1783.3bp,714.08bp)  .. (5);
  \draw [->] (12) ..controls (1588.2bp,1148.1bp) and (2026.0bp,1074.1bp)  .. (2026.0bp,955.0bp) .. controls (2026.0bp,955.0bp) and (2026.0bp,955.0bp)  .. (2026.0bp,665.0bp) .. controls (2026.0bp,458.33bp) and (2019.9bp,387.78bp)  .. (1905.0bp,216.0bp) .. controls (1880.7bp,179.67bp) and (1875.3bp,166.8bp)  .. (1838.0bp,144.0bp) .. controls (1790.7bp,115.08bp) and (1727.1bp,101.45bp)  .. (20);
  \draw [->] (13) ..controls (1431.8bp,225.77bp) and (1371.6bp,358.98bp)  .. (1318.0bp,468.0bp) .. controls (1294.1bp,516.52bp) and (1281.8bp,525.67bp)  .. (1262.0bp,576.0bp) .. controls (1224.9bp,670.52bp) and (1147.6bp,977.55bp)  .. (4);
  \draw [->] (22) ..controls (791.93bp,588.47bp) and (738.57bp,732.57bp)  .. (658.0bp,828.0bp) .. controls (646.12bp,842.08bp) and (629.81bp,854.58bp)  ..  (3);
  \draw [->] (18) ..controls (1571.0bp,652.28bp) and (1623.4bp,637.9bp)  .. (1662.0bp,612.0bp) .. controls (1846.4bp,488.22bp) and (1990.0bp,457.09bp)  .. (1990.0bp,235.0bp) .. controls (1990.0bp,235.0bp) and (1990.0bp,235.0bp)  .. (1990.0bp,161.0bp) .. controls (1990.0bp,43.814bp) and (1805.4bp,22.889bp)  ..  (19);
  \draw [->] (5) ..controls (1870.7bp,753.68bp) and (1883.0bp,749.53bp)  .. (1883.0bp,738.0bp) .. controls (1883.0bp,729.62bp) and (1876.5bp,725.14bp)  ..  (5);
  \draw [->] (8) ..controls (494.69bp,321.68bp) and (507.0bp,317.53bp)  .. (507.0bp,306.0bp) .. controls (507.0bp,297.62bp) and (500.5bp,293.14bp)  ..  (8);
  \draw [->] (26) ..controls (272.04bp,508.41bp) and (277.17bp,506.07bp)  .. (282.0bp,504.0bp) .. controls (317.07bp,488.96bp) and (357.94bp,473.4bp)  .. (10);
  \draw [->] (19) ..controls (1791.5bp,20.08bp) and (2008.0bp,34.089bp)  .. (2008.0bp,161.0bp) .. controls (2008.0bp,235.0bp) and (2008.0bp,235.0bp)  .. (2008.0bp,235.0bp) .. controls (2008.0bp,457.09bp) and (1864.4bp,488.22bp)  .. (1680.0bp,612.0bp) .. controls (1640.6bp,638.48bp) and (1586.6bp,652.91bp)  ..  (18);
  \draw [->] (2) ..controls (1361.8bp,1181.5bp) and (1411.0bp,1060.8bp)  .. (1411.0bp,955.0bp) .. controls (1411.0bp,955.0bp) and (1411.0bp,955.0bp)  .. (1411.0bp,881.0bp) .. controls (1411.0bp,611.19bp) and (1445.9bp,285.66bp)  .. (13);
  \draw [->] (12) ..controls (1490.1bp,1131.4bp) and (1536.0bp,1080.8bp)  .. (1536.0bp,1027.0bp) .. controls (1536.0bp,1027.0bp) and (1536.0bp,1027.0bp)  .. (1536.0bp,809.0bp) .. controls (1536.0bp,768.56bp) and (1527.6bp,722.33bp)  ..  (18);
  \draw [->] (4) ..controls (1052.8bp,1070.2bp) and (966.0bp,1024.9bp)  .. (966.0bp,955.0bp) .. controls (966.0bp,955.0bp) and (966.0bp,955.0bp)  .. (966.0bp,881.0bp) .. controls (966.0bp,738.34bp) and (940.12bp,697.89bp)  .. (866.0bp,576.0bp) .. controls (858.6bp,563.83bp) and (847.91bp,552.3bp)  ..  (22);
  \draw [->] (23) ..controls (543.71bp,572.63bp) and (492.0bp,657.52bp)  .. (492.0bp,737.0bp) .. controls (492.0bp,811.0bp) and (492.0bp,811.0bp)  .. (492.0bp,811.0bp) .. controls (492.0bp,853.74bp) and (472.57bp,900.22bp)  ..  (7);
  \draw [->] (8) ..controls (505.11bp,327.61bp) and (536.61bp,345.31bp)  ..  (9);
  \draw [->] (15) ..controls (1213.6bp,146.27bp) and (1510.3bp,108.78bp)  .. (20);
  \draw [->] (7) ..controls (533.58bp,953.08bp) and (686.62bp,947.39bp)  .. (803.0bp,900.0bp) .. controls (904.11bp,858.83bp) and (951.74bp,852.91bp)  .. (1002.0bp,756.0bp) .. controls (1034.9bp,692.47bp) and (1004.0bp,666.56bp)  .. (1004.0bp,595.0bp) .. controls (1004.0bp,595.0bp) and (1004.0bp,595.0bp)  .. (1004.0bp,521.0bp) .. controls (1004.0bp,433.25bp) and (995.86bp,411.61bp)  .. (991.0bp,324.0bp) .. controls (989.85bp,303.28bp) and (988.81bp,279.84bp)  ..  (11);
  \draw [->] (8) ..controls (574.96bp,289.94bp) and (850.43bp,253.21bp)  ..  (11);
  \draw [->] (7) ..controls (490.21bp,930.72bp) and (529.32bp,911.43bp)  ..  (3);
  \draw [->] (1) ..controls (1419.4bp,1308.7bp) and (1988.0bp,1273.7bp)  .. (1988.0bp,955.0bp) .. controls (1988.0bp,955.0bp) and (1988.0bp,955.0bp)  .. (1988.0bp,881.0bp) .. controls (1988.0bp,816.52bp) and (1914.0bp,772.57bp)  ..  (5);
  \draw [->] (14) ..controls (1072.8bp,242.53bp) and (1224.0bp,480.88bp)  .. (1408.0bp,612.0bp) .. controls (1431.4bp,628.66bp) and (1460.8bp,642.81bp)  .. (18);
  \draw [->] (17) ..controls (1216.8bp,776.43bp) and (1405.3bp,707.27bp)  ..  (18);
  \draw [->] (4) ..controls (1017.7bp,1080.8bp) and (805.74bp,1046.5bp)  ..  (6);
  \draw [->] (13) ..controls (1484.5bp,188.05bp) and (1498.4bp,202.65bp)  .. (1510.0bp,216.0bp) .. controls (1564.0bp,277.94bp) and (1579.9bp,291.95bp)  .. (1626.0bp,360.0bp) .. controls (1711.5bp,486.3bp) and (1794.8bp,649.53bp)  ..  (5);
  \draw [->] (11) ..controls (1121.5bp,219.57bp) and (1526.4bp,179.14bp)  ..  (16);
  \draw [->] (6) ..controls (717.13bp,987.45bp) and (752.03bp,943.2bp)  .. (772.0bp,900.0bp) .. controls (805.97bp,826.51bp) and (819.57bp,798.56bp)  .. (800.0bp,720.0bp) .. controls (764.96bp,579.36bp) and (737.68bp,545.07bp)  .. (647.0bp,432.0bp) .. controls (636.81bp,419.29bp) and (623.25bp,407.3bp)  ..  (9);
  \draw [->] (23) ..controls (661.17bp,462.84bp) and (883.42bp,307.43bp)  ..  (11);
  \draw [->] (25) ..controls (861.86bp,475.63bp) and (774.17bp,399.69bp)  .. (687.0bp,360.0bp) .. controls (624.31bp,331.46bp) and (544.55bp,317.36bp)  ..  (8);
  \draw [->] (22) ..controls (719.5bp,504.25bp) and (533.19bp,471.24bp)  ..  (10);
  \draw [->] (23) ..controls (580.04bp,479.67bp) and (582.23bp,435.21bp)  ..  (9);
  \draw [->] (19) ..controls (1780.3bp,26.29bp) and (1949.0bp,43.408bp)  .. (1996.0bp,72.0bp) .. controls (2038.5bp,97.875bp) and (2064.0bp,111.22bp)  .. (2064.0bp,161.0bp) .. controls (2064.0bp,379.0bp) and (2064.0bp,379.0bp)  .. (2064.0bp,379.0bp) .. controls (2064.0bp,527.06bp) and (2026.3bp,571.55bp)  .. (1930.0bp,684.0bp) .. controls (1914.3bp,702.3bp) and (1890.8bp,716.52bp)  ..  (5);
  \draw [->] (16) ..controls (1648.2bp,183.21bp) and (1620.1bp,198.92bp)  .. (1598.0bp,216.0bp) .. controls (1436.6bp,340.73bp) and (1434.4bp,415.27bp)  .. (1273.0bp,540.0bp) .. controls (1254.2bp,554.54bp) and (1231.0bp,568.09bp)  ..  (21);
  \draw [->] (5) ..controls (1762.6bp,725.45bp) and (1654.5bp,705.83bp)  .. (1565.0bp,684.0bp) .. controls (1559.1bp,682.56bp) and (1552.9bp,680.92bp)  ..  (18);
  \draw [->] (21) ..controls (1030.6bp,580.66bp) and (500.72bp,537.79bp)  ..  (24);
  \draw [->] (22) ..controls (805.43bp,610.78bp) and (778.01bp,874.07bp)  .. (752.0bp,900.0bp) .. controls (714.58bp,937.3bp) and (555.44bp,948.54bp)  ..  (7);
\end{tikzpicture}}
};
\end{tikzpicture}
\caption{Three implementations of the TBURST4 component of the AMBA
  bus controller~\cite{FK:2016}. Game-based synthesis produces a Mealy machine with the shape shown on the
  left with 14 states and 61 cycles. Bounded synthesis produces the
  implementation in the middle with 7 states and 19 cycles. The implementation on the
  right, produced by bounded cycle synthesis, has 7 states and 7 cycles, which is the
  minimum.}
\label{fig:amba}
\end{figure}

A direct comparison of the three algorithms is shown in
Fig.~\ref{fig:amba}. The figure depicts the shape of the synthesized
implementations from the AMBA TBURST4 specification~\cite{FK:2016} using, from left to right, game-based
synthesis, bounded synthesis, and bounded cycle synthesis.
Even just based on a superficial visual comparison, it is immediately clear that the output-sensitive
algorithms produce dramatically simpler implementations.


\paragraph{Acknowledgement.}
This work was partially supported by the European Research Council (ERC) Grant OSARES (No.\ 683300).

\bibliographystyle{abbrv}
\bibliography{bib}

\end{document}